\documentclass[journal]{IEEEtran}
\usepackage{amsmath}
\usepackage{amssymb}
\usepackage[mathscr]{eucal}
\usepackage{mathtools}
\usepackage[pdftex]{graphicx}
\usepackage{xcolor}
\usepackage{cite}
\usepackage{url}
\usepackage{soul}

\newtheorem{definition}{\textbf{Definition}}
\newtheorem{proposition}{\textbf{Proposition}}

\begin{document}

\title{Multi-target Detection with \\an Arbitrary Spacing Distribution}

\author{Ti-Yen Lan,
	  Tamir Bendory,
	  Nicolas Boumal
	  and~Amit Singer
\thanks{
T.-Y.\ Lan, T.\ Bendory, N.\ Boumal and A.\ Singer are with the Program in Applied and Computational Mathematics and the Mathematics Department, Princeton University, Princeton, NJ 08544, USA (e-mail: tiyenlan@princeton.edu; tamir.bendory@princeton.edu; nboumal@math.princeton.edu; amits@math.princeton.edu).
The research is supported in parts by Award Number R01GM090200 from the NIGMS, FA9550-17-1-0291 from AFOSR, Simons Foundation Math+X Investigator Award, the Moore Foundation Data-Driven Discovery Investigator Award, and NSF BIGDATA Award IIS-1837992.
NB is partially supported by NSF award DMS-1719558.}
}

\maketitle

\begin{abstract}
Motivated by the structure reconstruction problem in single-particle cryo-electron microscopy, we consider the multi-target detection model, where multiple copies of a target signal occur at unknown locations in a long measurement, further corrupted by additive Gaussian noise.
At low noise levels, one can easily detect the signal occurrences and estimate the signal by averaging.
However, in the presence of high noise, which is the focus of this paper, detection is impossible.
Here, we propose two approaches---autocorrelation analysis and an approximate expectation maximization algorithm---to reconstruct the signal without the need to detect signal occurrences in the measurement.
In particular, our methods apply to an arbitrary spacing distribution of signal occurrences.
We demonstrate reconstructions with synthetic data and empirically show that the sample complexity of both methods scales as SNR${}^{-3}$ in the low SNR regime.
\end{abstract}

\begin{IEEEkeywords}
autocorrelation analysis, expectation maximization, frequency marching, cryo-EM, blind deconvolution.
\end{IEEEkeywords}

\IEEEpeerreviewmaketitle

\section{Introduction}
\begin{figure}[h]
\includegraphics[scale=0.66]{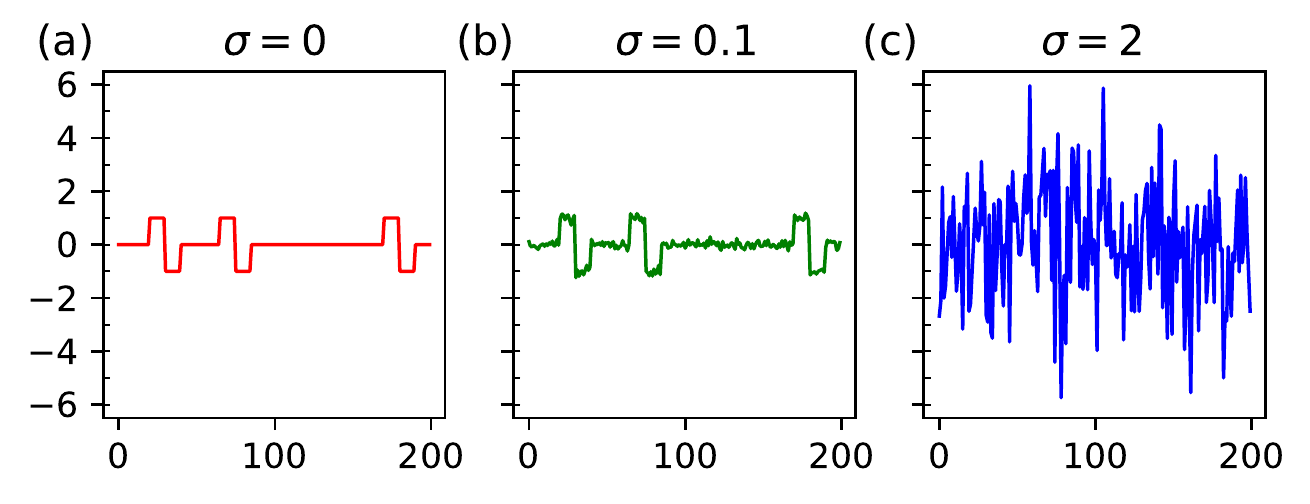}
\caption{An example of a measurement in the MTD model corrupted by additive Gaussian noise with (a) $\sigma = 0$, (b) $\sigma = 0.1$ (high SNR regime) and (c) $\sigma = 2$ (low SNR regime).}
\label{fig:mtd}
\end{figure}

We consider the multi-target detection (MTD) problem~\cite{Bendory19} to estimate a signal $x \in \mathbb{R}^L$ from a long, noisy measurement
\begin{equation}
\label{MTD}
y = s \ast x + \varepsilon,\quad \varepsilon \sim \mathcal{N}(0, \sigma^2 I_N),
\end{equation}
where $y \in \mathbb{R}^N$ is the linear convolution of an unknown binary sequence $s \in \{0, 1\}^{N-L+1}$ with the signal, further corrupted by additive white Gaussian noise with zero mean and variance $\sigma^2$, and we assume $L \ll N$.
Both $x$ and $s$ are treated as deterministic variables.
The signal length $L$ and the noise variance $\sigma^2$ are assumed to be known.
The non-zero entries of $s$ indicate the starting positions of the signal occurrences in $y$.
We require the signal occurrences not to overlap, so consecutive non-zero entries of $s$ are separated by at least $L$ positions.
Figure~\ref{fig:mtd} gives an example of the measurement $y$ that contains three signal occurrences at different noise levels.

MTD belongs to the wider class of blind deconvolution problems~\cite{Kondur96}, which have applications in astronomy~\cite{Jefferies93,Schulz93}, microscopy~\cite{Sarder06,McNally99}, system identification~\cite{AM97}, and motion deblurring~\cite{Levin06,Cho09}, among others.
The main difference is that we are only interested in estimating the signal $x$: we treat $s$ as a \emph{nuisance variable}, while most of the literature on blind deconvolution aims to estimate both $x$ and $s$.
This distinction allows us to estimate the signal at higher noise levels, which are usually not addressed in the literature on blind deconvolution, specifically because $s$ \emph{cannot} be accurately estimated in such regimes.
We give a theoretical argument for this claim below, and corroborate it with numerical experiments in Section~\ref{section:numerical}.

This high-noise MTD model has been studied in~\cite{Bendory18b,Bendory19} under the assumption that the signal occurrences either are well separated or follow a Poisson distribution.
In applications, however, the signal occurrences in the measurement might be arbitrarily close to each other.
In this study, we extend the framework of~\cite{Bendory18b,Bendory19} to allow an arbitrary spacing distribution of signal occurrences by simultaneously estimating the signal and the \emph{distance distribution} between consecutive signal occurrences.

The solution of the MTD problem is straightforward at high signal-to-noise ratio (SNR), such as the case shown in Figure~\ref{fig:mtd}(b).
The signal can be estimated by first detecting the signal occurrences and then averaging.
In the low SNR regime however, this simple method becomes problematic due to the difficulty of detection, as illustrated by Figure~\ref{fig:mtd}(c).
More than difficulty, reliable detection becomes impossible beyond a critical noise level. This can be understood from~\cite[Proposition 3.1]{Bendory18b}, which we reproduce here (the proof is based on the Neyman--Pearson lemma).
\begin{proposition}
\label{prop1}
Consider two known vectors $\theta_0 = x$ and $\theta_1 = 0$ in $\mathbb{R}^L$ and a random variable $\eta$ taking value 0 or 1 with equal probability. 
We observe the random vector $X \in \mathbb{R}^L$ with the following distribution:
\begin{align*}
X \sim 
\begin{cases}
\mathcal{N}(\theta_0, \sigma^2 I_L)\quad\mathrm{if}~\eta = 0 \\
\mathcal{N}(\theta_1, \sigma^2 I_L)\quad\mathrm{if}~\eta = 1.
\end{cases}
\end{align*}
For any deterministic estimator $\hat{\eta}$ of $\eta$,
\begin{align*}
\lim_{\sigma\rightarrow\infty} p(\hat{\eta} = \eta) \leq \frac{1}{2},
\end{align*}
that is, the probability of success for the best deterministic estimator of $\eta$ is no better than a random guess in the limit of $\sigma\rightarrow\infty$.
\end{proposition}

In our context, this simple fact can be interpreted as follows: even if we know the true signal $x$ and an oracle provides windows of length $L$ in the observation $y$ which, with equal probability, contain either the noisy signal or pure noise, deciding whether a given window contains one or the other cannot be done significantly better than a random chance decision at high noise level.
Results similar in spirit (but distinct in mathematical tools) are derived in~\cite{Aguerrebere16}.

In this work, we suggest to circumvent the estimation of $s$ by estimating the signal $x$ directly.
Specifically, we propose two different approaches---autocorrelation analysis and an approximate version of the expectation maximization (EM) algorithm~\cite{Dempster77}.
Since these methods do not require estimating~$s$,~Proposition~\ref{prop1} does not limit their performance in the low SNR regime.

Autocorrelation analysis relates the autocorrelations calculated from the noisy measurement to the signal.
The signal is then estimated by fitting the autocorrelations through least squares.
This approach is efficient as it requires only one pass over the data to calculate the autocorrelations; this is of particular importance as the data size grows.

In the second approach, the approximate EM algorithm, the signal is reconstructed by iteratively maximizing the data likelihood, which marginalizes over $s$; importantly, it does not estimate it explicitly.
In contrast to autocorrelation analysis, the approximate EM algorithm scans through the whole dataset in each iteration, and hence requires much longer computational time.

In this study, we demonstrate the reconstruction of the underlying signal from the noisy measurement using the two proposed approaches.
Our numerical experiments show that the approximate EM algorithm provides slightly more accurate estimates of the signal in the low SNR regime, whereas autocorrelation analysis is considerably faster, especially at low SNR.
It is empirically shown that the sample complexity of both approaches scales as SNR${}^{-3}$ at low SNR, with details discussed later in the text.
In the high SNR regime, the sample complexity of both methods scales as SNR${}^{-1}$, the same as the sample complexity of the simple method that estimates the signal by first detecting the signal occurrences and then averaging.

The main contributions of this work are as follows.
\begin{enumerate}
\item We formalize MTD for an arbitrary signal spacing distribution, making it a more realistic model for applications.
\item We propose two algorithms to solve the MTD problem.
In particular, our algorithm based on autocorrelations illustrates why, to recover $x$, we need not estimate all of $s$, but rather only a concise summary of it; and why it is possible, in principle at least, to solve this problem for arbitrary noise levels (given a sufficiently long observation).
For our second algorithm, we note that the popular EM method is intractable, but we show how to implement an approximation of it, which performs well in practice.
\item For both algorithms, we design a coarse-to-fine multi-resolution scheme to alleviate issues pertaining to non-convexity.
This is related to the ideas of frequency marching which are often used in cryo-electron microscopy (cryo-EM)~\cite{Barnett17,Scheres12}.
\end{enumerate}

\section{Autocorrelation analysis}
\label{section:AA}
In what follows, we discuss autocorrelations of both the signal $x$ (of length $L$) and the measurement $y$ (of length $N$).
To keep notation general, we here consider a sequence $z$ of length $m$, and define its autocorrelations of order $q = 1, 2, \dots$ for any integer shifts $l_1, l_2, \dots, l_{q-1}$ as
\begin{equation}
a_z^q[l_1, l_2, \dots, l_{q-1}] = \frac{1}{m} \sum_{i=0}^{m-1} z[i] z[i+l_1] \cdots z[i+l_{q-1}], \nonumber
\end{equation}
where $z$ is zero-padded for indices out of the range $[0, m-1]$.
We have $m = L$ when $z$ represents the signal $x$ and $m = N$ when $z$ represents the measurement $y$.
Since the autocorrelations only depend on the differences of the integer shifts and are invariant under any permutation of the shifts, for the second- and third-order autocorrelations we have the symmetries
\begin{align}
\label{symmetries}
a_z^2[l] = a_z^2[-l],~a_z^3[l_1, l_2] = a_z^3[l_2, l_1] = a_z^3[-l_1, l_2-l_1].
\end{align}
For applications of higher-order autocorrelations and their symmetries, see for example~\cite{Giannakis89,Swami90,Sadler92,Aizenbud19}.

In this section, we describe how the autocorrelations of the noisy measurement $y$ are related to the underlying signal~$x$.
These relations are later used to estimate $x$ without the need to identify the locations of signal occurrences, which are nuisance variables and difficult, if not impossible, to determine reliably at high noise levels.
For completeness, we include a brief discussion of the special case where the signals are well separated and refer the reader to \cite{Bendory18b,Bendory19} for details.
The generalization to an arbitrary signal spacing distribution follows.

\subsection{Well-separated signals}
\label{section:AA_ws}
The signals are said to be well-separated when the consecutive non-zero entries of $s$ are separated by at least $2L - 1$ positions.
Under this condition, the autocorrelations of $y$ with integer shifts $l_1, l_2, \dots, l_{q-1}$ within the range $[-(L-1), L-1]$ are unaffected by the relative positions of signal occurrences.
As a result, these autocorrelations of $y$ provide fairly direct information about those of $x$, and therefore about $x$ itself.

Due to the presence of Gaussian noise, the entries of $y$ are stochastic.
Taking the expectations of the first three autocorrelations of $y$ with respect to the distribution of Gaussian noise, we obtain these relations (see Appendix~\ref{aa_relations}):
\begin{align}
\label{a1_ws}
\mathop{\mathbb{E}_\varepsilon} \{a_y^1\} &= \rho_0 a_x^1 \\
\label{a2_ws}
\mathop{\mathbb{E}_\varepsilon} \{a_y^2[l]\} &= \rho_0 a_x^2[l] + \sigma^2 \delta[l] \\
\label{a3_ws}
\mathop{\mathbb{E}_\varepsilon} \{a_y^3[l_1, l_2]\} &= \rho_0 a_x^3[l_1, l_2] \nonumber \\
	& \quad + \rho_0 a_x^1 \sigma^2 (\delta[l_1] + \delta[l_2] + \delta[l_1-l_2]),
\end{align}
where $0 \leq l < L$ and $0 \leq l_1 \leq l_2 < L$.
Here, $\rho_0 = ML/N$ denotes the signal density, where $M$ is the number of signal occurrences in $y$. 
We assume $M$ grows with $N$ at the constant rate $\rho_0/L$.
The delta functions, defined by $\delta[0] = 1$ and $\delta[l\neq0] = 0$, are due to the autocorrelations of the white Gaussian noise.
As indicated by the studies in phase retrieval~\cite{Bruck79,Bates82,Beinert15}, in general a 1D signal $x$ cannot be uniquely determined by its first two autocorrelations.
It is thus necessary (and generically sufficient~\cite{Bendory19}) to include the third-order autocorrelations to uniquely determine $x$.

The expectations of the autocorrelations of $y$ are estimated by averaging over the given noisy measurement.
This data reduction requires only one pass over the data, which is a great computational advantage as the data size grows.
As shown in Appendix~\ref{aa_relations}, for a given signal $x$, the average over the noisy entries of $y$ gives the relations:
\begin{align}
a_y^1 &= \mathop{\mathbb{E}_\varepsilon} \{a_y^1\} + \eta^1 \nonumber \\
a_y^2[l] &= \mathop{\mathbb{E}_\varepsilon} \{a_y^2[l]\} + \eta^2[l] \nonumber \\
a_y^3[l_1, l_2] &= \mathop{\mathbb{E}_\varepsilon} \{a_y^3[l_1, l_2]\} + \eta^3[l_1, l_2], \nonumber
\end{align}
where $\eta^1, \eta^2[l]$ and $\eta^3[l_1, l_2]$ are random variables with zero mean and variances $\mathcal{O}\big(\sigma^2/N\big)$, $\mathcal{O}\big((\sigma^2 + \sigma^4)/N\big)$ and $\mathcal{O}\big((\sigma^2 + \sigma^6)/N\big)$ respectively.
For autocorrelations of order $q$, the standard deviations scale as $\sigma^{q}/\sqrt{N}$ at high noise levels.
Therefore, we need $\sqrt{N}/\sigma^{q} \gg 1$ in order for the $a_y^q$ calculated from the noisy measurement to be a good estimator for $\mathop{\mathbb{E}_\varepsilon} \{a_y^q\}$, and thus to establish a reliable relation with the signal~$x$ such as \eqref{a1_ws}-\eqref{a3_ws}.
Since the SNR is proportional to $\sigma^{-2}$, the sample complexity therefore scales as SNR${}^{-q}$.
We also expect the error of the reconstructed signal to depend on the errors of the highest-order autocorrelations used in the analysis at high noise levels.

We estimate the signal density $\rho_0$ and signal $x$ by fitting the first three autocorrelations of $y$ via non-linear least squares:
\begingroup
\allowdisplaybreaks
\begin{align}
\label{cost_ws}
f(\hat{\rho}_0, \hat{x}) &= (a_y^1 - \hat{\rho}_0 a_{\hat{x}}^1)^2 + \frac{1}{L} \sum_{l=0}^{L-1} (a_y^2[l] - \hat{\rho}_0 a_{\hat{x}}^2[l] - \sigma^2\delta[l])^2 \nonumber \\
&+ \frac{2}{L(L+1)} \sum_{l_2=0}^{L-1} \sum_{l_1=0}^{l_2} \bigg( a_y^3[l_1, l_2] - \hat{\rho}_0 a_{\hat{x}}^3[l_1, l_2] \nonumber \\
&\hspace{3em} - \hat{\rho}_0 a_{\hat{x}}^1 \sigma^2 (\delta[l_1] + \delta[l_2] + \delta[l_1-l_2]) \bigg)^2.
\end{align}
\endgroup
The weights are chosen as the inverse of the number of terms in the respective sums.
Since the autocorrelations have symmetries, as indicated in \eqref{symmetries}, the summations above are restricted to the non-redundant shifts.
Due to the errors in estimating $\mathop{\mathbb{E}_\varepsilon} \{a_y^3\}$ with $a_y^3$, we expect the root-mean-square error of the reconstructed signal $\hat{x}$,
\begin{equation}
\mathrm{RMSE}(\hat{x}) = \frac{||\hat{x} - x||_2}{||x||_2}, \nonumber
\end{equation}
to scale as $\sigma^3$ in the low SNR regime.

We mention that there exist alternative methods to fit the observed autocorrelations.
One possibility is to reformulate the problem as a tensor sensing problem: the autocorrelations are then linear functions of an $(L+1)\times(L+1)\times(L+1)$ tensor.
Such an approach has been proven to be effective for the related problems of multi-reference alignment~\cite{Bandeira17,Perry19} and phase retrieval~\cite{Candes13}.
Similar formulations are also found in tensor optimization~\cite{Auria13,Holger15}.
Nevertheless, this method requires lifting the original problem to a higher-dimensional space, which increases the computational burden and scales poorly with the dimension of the problem. 
In contrast, our least-squares formulation operates in the ambient dimension of the problem and thus might be applicable to high-dimensional setups.

\subsection{Arbitrary spacing distribution}
\label{section:AA_asd}
The condition of well-separated signals can be further relaxed to allow arbitrary spacing distribution by assuming that the signal occurrences in any subset of $y$ follow the same spatial distribution.
To this end, we define the pair separation function as follows.
\begin{definition}
\label{def:psf}
For a given binary sequence $s$ identifying $M$ starting positions of signal occurrences (that is, $\sum_i s[i] = M$), the \emph{pair separation function} $\xi[l]$ is defined as
\begin{equation}
\xi[l] = \frac{1}{M-1} \sum_{k=1}^{M-1} \delta[s_{k-1} + l - s_k], 
\end{equation}
where $s_k$ indicates the index of the $k^{\mathrm{th}}$ non-zero entry of the sequence $s$.
It is not hard to see that $\sum_l \xi[l] = 1$.
\end{definition}

In particular, we force $\xi[0] = \cdots = \xi[L-1] = 0$ since we exclude overlapping occurrences; then $\xi[L]$ is the fraction of pairs of consecutive signal occurrences that occur right next to each other (no spacing at all), $\xi[L+1]$ is the fraction of pairs of consecutive signal occurrences that occur with one signal-free entry in between them, etc.

In contrast to the well-separated model, autocorrelations of $y$ may now involve correlating distinct occurrences of $x$, which may be in various relative positions.
The crucial observation is that these autocorrelations depend only indirectly on $s$, namely, through $L-1$ entries of the unknown $\xi$, which have much smaller dimensions than $s$.
As shown in Appendix A, for a given signal $x$, the autocorrelations of the given measurement $y$ gives the relations:
\begingroup
\allowdisplaybreaks
\begin{align}
\label{a1_asd}
a_y^1 &= \mathop{\mathbb{E}_\varepsilon} \{a_y^1\} + \eta_1' = \rho_0 a_x^1 + \eta_1' \\
\label{a2_asd}
a_y^2[l] &= \mathop{\mathbb{E}_\varepsilon} \{a_y^2[l]\} + \eta_2'[l] \nonumber \\
	&= \rho_0 a_x^2[l] + \rho_0 \hspace{-3pt} \sum_{j=L}^{L+l-1} \xi[j] a_x^2[j-l] + \sigma^2 \delta[l] + \eta_2'[l] \\
\label{a3_asd}
a_y^3[l_1, &l_2] = \mathop{\mathbb{E}_\varepsilon} \{a_y^3[l_1, l_2]\} + \eta_3'[l_1, l_2] \nonumber \\
	&\hspace{10.5pt}= \rho_0 a_x^3[l_1, l_2] 
		+ \rho_0 \hspace{-7pt} \sum_{j = L}^{L+l_2-l_1-1} \hspace{-0.75em} \xi[j] a_x^3[j-l_2, j+l_1-l_2] \nonumber \\
	&\hspace{10.5pt}+ \rho_0 \hspace{-3pt} \sum_{j = L}^{L+l_1-1} \xi[j] a_x^3[l_2-l_1, j-l_1] 
		+ \rho_0 a_x^1 \sigma^2 \big( \delta[l_1] \nonumber \\
	&\hspace{10.5pt}+ \delta[l_2] + \delta[l_1-l_2] \big) + \eta_3'[l_1, l_2],
\end{align}
\endgroup
where $0 \leq l < L$ and $0 \leq l_1 \leq l_2 < L$.
The random variables $\eta_1'$, $\eta_2'[l]$ and $\eta_3'[l_1, l_2]$ have zero mean and variances $\mathcal{O}\big(\sigma^2/N\big)$, $\mathcal{O}\big((\sigma^2 + \sigma^4)/N\big)$ and $\mathcal{O}\big((\sigma^2 + \sigma^6)/N\big)$ respectively.
Note that \eqref{a1_asd}--\eqref{a3_asd} reduce to \eqref{a1_ws}--\eqref{a3_ws} when $\xi[L] = \xi[L+1] = \cdots = \xi[2L-2] = 0$, as required by the condition of well-separated signals.
Expressions \eqref{a1_asd}--\eqref{a3_asd} further simplify upon defining
\begin{align}
\label{rho1}
\rho_1[i] = \rho_0 \xi[i+L], \quad i = 0, 1, \dots, L-2.
\end{align}

After calculating the first three autocorrelations of $y$ from the noisy measurement, we estimate the signal $x$ and the parameters $\rho_0$ and $\rho_1$ by fitting the autocorrelations of $y$ through the non-linear least squares
\begingroup
\allowdisplaybreaks
\begin{align}
\label{cost_asd}
&f(\hat{x}, \hat{\rho}_0, \hat{\rho}_1) = (a_y^1 - \hat{\rho}_0 a_{\hat{x}}^1)^2 \nonumber \\
	&\hspace{-0.5em}+ \frac{1}{L} \sum_{l=0}^{L-1} \bigg( a_y^2[l] - \hat{\rho}_0 a_{\hat{x}}^2[l] - \sigma^2\delta[l]
		-\hspace{-0.5em} \sum_{j=L}^{L+l-1} \hspace{-0.25em} \hat{\rho}_1[j-L] a_{\hat{x}}^2[j-l] \bigg)^2  \nonumber \\
	&\hspace{-0.5em}+ \frac{2}{L(L+1)} \sum_{l_2=0}^{L-1} \sum_{l_1=0}^{l_2} \bigg( a_y^3[l_1, l_2] 
		- \hat{\rho}_0 a_{\hat{x}}^3[l_1, l_2] - \hat{\rho}_0 a_{\hat{x}}^1 \sigma^2 \big( \delta[l_1] \nonumber \\
	&\hspace{-0.5em}+ \delta[l_2] + \delta[l_1-l_2] \big) 
		-\hspace{-0.5em} \sum_{j = L}^{L+l_2-l_1-1} \hspace{-0.75em} 
		\hat{\rho}_1[j-L] a_{\hat{x}}^3[j-l_2, j+l_1-l_2] \nonumber \\
	&\hspace{-0.5em}- \sum_{j = L}^{L+l_1-1} \hat{\rho}_1[j-L] a_{\hat{x}}^3[l_2-l_1, j-l_1] \bigg)^2.
\end{align}
\endgroup
Numerically, we find that this problem can often be solved, meaning that, even though $s$ cannot be estimated, we can still estimate $x$ and the summarizing statistics $\rho_0$ and $\rho_1$.
As discussed in Section~\ref{section:AA_ws}, the RMSE of the reconstructed signal is expected to scale as $\sigma^3$ in the low SNR regime owing to the errors in estimating $\mathop{\mathbb{E}_\varepsilon} \{a_y^3\}$ with $a_y^3$.

\subsection{Frequency marching}
To minimize the least squares in (\ref{cost_ws}) and (\ref{cost_asd}), we use the trust-regions method in Manopt~\cite{Boumal14} over the product of the Euclidean manifold with the constraints that $\hat{\rho}_0$ and $\hat{\rho}_1$ are positive.
However, as the least squares problems are inherently non-convex, we observe that the iterates of the trust-regions method used for minimization are liable to stagnate in local minima.
To alleviate this issue, we adopt the frequency marching scheme~\cite{Barnett17} in our optimization.
The idea behind frequency marching is based on the following heuristics:
\begin{enumerate}
\item The coarse-grained (low-resolution) version of the original problem has a smoother optimization landscape so that it is empirically more likely for the iterates of the optimization algorithm to reach the global optimum of the coarse-grained problem.
\item Intuitively, the global optimum of the original problem can be reached more easily by following the path along the global optima of a series of coarse-grained problems with incremental resolution.
\end{enumerate}
Our goal is to guide the iterates of the optimization algorithm to reach the global optimum of the original problem by successively solving the coarse-grained problems, which are warm-started with the solution from the previous stage.

The coarse-grained problems are characterized by the order of the Fourier series, $n_{\max} = 1, 2, \dots, \lfloor L/2 \rfloor$, used to express the low-resolution approximate $x^{(n_{\max})}$ by
\begin{equation}
\label{fourier_expansion}
x^{(n_{\max})}[l] = c_0 + \sum_{n=1}^{n_{\max}} c_n \cos \bigg( \frac{2\pi nl}{L} \bigg) + d_n \sin \bigg( \frac{2\pi nl}{L} \bigg),\hspace{-0.15em} 
\end{equation}
where $l = 0, 1, \dots, L-1$.
Instead of the entries of $x$, the least squares are minimized with respect to the Fourier coefficients in our frequency marching scheme.
The order $n_{\max}$ is related to the spatial resolution by Nyquist rate:
\begin{equation}
\Delta x =
\begin{cases}
	1 & \text{if}~n_{\max} = (L-1)/2, \\
	L/2n_{\max} & \text{otherwise.} \nonumber
\end{cases}
\end{equation}
The spatial resolution sub\-divides the signal $x$ into $L' = L/\Delta x$ units and represents the ``step size'' of the shifts for the coarse-grained autocorrelations.

We define the coarse-grained autocorrelations of $x^{(n_{\max})}$ of order $q$ for integer shifts $l_1, l_2, \dots, l_{q-1}$ as
\begin{align}
&b_{x^{{(n_{\max})}}}^q [l_1, l_2, \dots, l_{q-1}] \nonumber \\
	&\hspace{-0.3em}= a_{x^{{(n_{\max})}}}^q[\lfloor l_1 \Delta x \rceil, \lfloor l_2 \Delta x \rceil, \dots, \lfloor l_{q-1} \Delta x \rceil], \nonumber
\end{align}
where $\lfloor \cdot \rceil$ rounds the argument to the nearest integer.
The coarse-grained autocorrelations of $y$ are given by sub-sampling the original autocorrelations calculated from the full measurement.
With $b[l]$ denoting the bin centered at $l\Delta x$, where $l = 0, 1, \dots, L'-1$, we estimate the coarse-grained autocorrelations of $y$ by
\begin{align}
b_y^1 &= a_y^1 \nonumber \\
b_y^2[l] &= B_2^{-1} \sum_{i \in b[l]} \bigg( a_y^2[i] - \sigma^2 \delta[i] \bigg) \nonumber \\
b_y^3[l_1, l_2] &= B_3^{-1}
			\sum_{\substack{i_1 \in b[l_1] \\ i_2 \in b[l_2] \\ i_1 \leq i_2}} \bigg( a_y^3[i_1, i_2] \nonumber \\
			&\hspace{3em} - a_y^1 \sigma^2 \big( \delta[i_1] + \delta[i_2] + \delta[i_1-i_2] \big) \bigg), \nonumber
\end{align}
where $0 \leq l < L'$, $0 \leq l_1 \leq l_2 < L'$, and $B_2$ and $B_3$ represent the number of terms in the respective sums.

Following the discussion in Section~\ref{section:AA_asd}, we relate the first three autocorrelations of $y$ to those of $x^{{(n_{\max})}}$, as defined in \eqref{fourier_expansion}, by
\begingroup
\allowdisplaybreaks
\begin{align}
b_y^1\ &\approx \rho_0 b_{x^{{(n_{\max})}}}^1 \nonumber \\
b_y^2[l] &\approx \rho_0 b_{x^{{(n_{\max})}}}^2[l] 
	+ \hspace{-0.5em} \sum_{j=L'}^{L'+l-1} \rho_1^{(n_{\max})}[j-L'] b_{x^{{(n_{\max})}}}^2[j-l] \nonumber \\
b_y^3[l_1, l_2]\} &\approx \rho_0 b_{x^{{(n_{\max})}}}^3[l_1, l_2] \nonumber \\
	&\hspace{-1em}+ \sum_{j = L'}^{L'+l_2-l_1-1} \hspace{-0.75em} \rho_1^{(n_{\max})}[j-L'] b_{x^{{(n_{\max})}}}^3[j-l_2, j+l_1-l_2] \nonumber \\
	&\hspace{-1em}+ \sum_{j = L'}^{L'+l_1-1} \rho_1^{(n_{\max})}[j-L'] b_{x^{{(n_{\max})}}}^3[l_2-l_1, j-l_1]. \nonumber
\end{align}
\endgroup
The autocorrelations are related by approximation instead of equality to reflect the errors due to the low-resolution approximation.
Above, we define $\rho_1^{(n_{\max})}$ as the product of the signal density $\rho_0$ and the coarse-grained pair separation function:
\begin{align}
\label{rho1_coarse}
\rho_1^{(n_{\max})}[i] = \rho_0 \xi^{(n_{\max})}[i+L'],\quad i = 0, 1, \dots, L'-2,
\end{align}
where $\xi^{(n_{\max})}$ is defined as
\begin{align}
\label{coarse_psf}
\xi^{(n_{\max})}[i] = \sum_{l=\max\{0, \lfloor (i - 1/2) \Delta x \rceil \}}^{\lfloor (i + 1/2) \Delta x \rceil - 1} \xi[l].
\end{align}
In each stage of frequency marching, we estimate the Fourier coefficients, which are related to $x^{(n_{\max})}$ through \eqref{fourier_expansion}, and the parameters $\rho_0, \rho_1^{(n_{\max})}$ by fitting the coarse-grained autocorrelations of $y$ through the non-linear least squares
\begin{align}
&\hspace{-1em}f(\{\hat{c}_0,\dots,\hat{c}_{n_{\max}}\}, 
	\{\hat{d}_1,\dots,\hat{d}_{n_{\max}}\}, \hat{\rho}_0, \hat{\rho}^{(n_{\max})}_1) \nonumber \\
&\hspace{-1.25em}= (b_y^1 - \hat{\rho}_0 b_{\hat{x}^{(n_{\max})}}^1)^2 
	+ \frac{1}{L'} \sum_{l=0}^{L'-1} \bigg( b_y^2[l] - \hat{\rho}_0 b_{\hat{x}^{(n_{\max})}}^2[l] \nonumber \\
&\hspace{-1.25em}- \sum_{j=L'}^{L'+l-1} \hat{\rho}_1^{(n_{\max})}[j-L'] b_{\hat{x}^{(n_{\max})}}^2[j-l] \bigg)^2 \nonumber \\
&\hspace{-1.25em}+ \frac{2}{L'(L'+1)} \sum_{l_2=0}^{L'-1} \sum_{l_1=0}^{l_2} 
	\bigg( b_y^3[l_1, l_2] - \hat{\rho}_0 b_{\hat{x}^{(n_{\max})}}^3[l_1, l_2] \nonumber \\
&\hspace{-1.25em}- \sum_{j = L'}^{L'+l_2-l_1-1} \hspace{-0.75em} 
	\hat{\rho}_1^{(n_{\max})}[j-L'] b_{\hat{x}^{(n_{\max})}}^3[j-l_2, j+l_1-l_2] \nonumber \\
&\hspace{-1.25em}- \sum_{j = L'}^{L'+l_1-1} \hat{\rho}_1^{(n_{\max})}[j-L'] 
	b_{\hat{x}^{(n_{\max})}}^3[l_2-l_1, j-l_1] \bigg)^2. \nonumber
\end{align}
Our frequency marching scheme increments the order $n_{\max}$ from $1$ to $\lfloor L/2 \rfloor$, and the computed solution of each stage is used to initialize optimization in the next stage.

\section{Expectation maximization}
\label{section:EM}
In this section, as an alternative to the autocorrelations approach, we describe an approximate EM algorithm to address both the cases of well-separated signals and arbitrary spacing distribution.
A frequency marching scheme is also designed to help the iterates of the EM algorithm converge to the global maximum of the data likelihood.

\subsection{Well-separated signals}
Given the measurement $y$ that follows the MTD model \eqref{MTD}, the maximum marginal likelihood estimator (MMLE) for the signal $x$ is the maximizer of the likelihood function $p(y | x)$.
Within the EM framework~\cite{Dempster77}, the nuisance variable $s$ is treated as a random variable drawn from some distribution under the condition of non-overlapping signal occurrences.
The EM algorithm estimates the MMLE by iteratively applying the expectation (E) and maximization (M) steps.
Specifically, given the current signal estimate $x_k$, the E-step constructs the expected log-likelihood function
\begin{align}
Q(x | x_k) = \sum_s p(s | y, x_k) \log p(y, s | x), \nonumber
\end{align}
where the summation runs over all admissible configurations of the binary sequence $s$.
The signal estimate is then updated in the M-step by maximizing $Q(x | x_k)$ with respect to $x$.
The major drawback of this approach is that the number of admissible configurations for $s$ grows exponentially with the problem size.
Therefore, the direct application of the EM algorithm is computationally intractable, even for very short measurements.

In our framework of the approximate EM algorithm, we first partition the measurement $y$ into $N_d = N/L$ non-overlapping segments, each of length $L$, and denote the $m^\mathrm{th}$ segment by~$y_m$.
Overall, the signal can occur in $2L - 1$ different ways when it is present in a segment.
The signal is estimated by the maximizer of the approximate likelihood function
\begin{align}
\label{data_likelihood}
p(y_0, y_1, \dots, y_{N_d-1} | x) \approx \prod_{m=0}^{N_d-1} p(y_m | x),
\end{align}
where we ignore the dependencies between segments.
Our approximate EM algorithm works by applying the EM algorithm to estimate the MMLE of \eqref{data_likelihood}, without any prior on the signal.
As we will see in Section~\ref{section:numerical}, the validity of the approximation is corroborated by the results of our numerical experiments.

Depending on the position of signal occurrences, the segment $y_m$ can be modeled by
\begin{equation}
y_m = C R_{l_m} Z x + \varepsilon_m,\quad \varepsilon_m \sim \mathcal{N}(0, \sigma^2 I_L). \nonumber
\end{equation}
Here, $Z$ first zero-pads $L$ entries to the left of $x$, and $R_{l_m}$ circularly shifts the zero-padded sequence by $l_m$ positions, that is,
\begin{equation}
(R_{l_m} Zx)[l] = (Zx)[(l+l_m)~\mathrm{mod}~2L], \nonumber
\end{equation}
where $l_m = 0, 1, \dots, 2L-1$ and is treated as a random variable.
The operator $C$ then crops the first $L$ entries of the circularly shifted sequence, which are further corrupted by additive white Gaussian noise.
In this generative model, $l_m = 0$ represents no signal occurrence in $y_m$, and $l_m = 1, \dots, 2L-1$ enumerate the $2L-1$ different ways a signal can appear in a segment.

In the E-step, our algorithm constructs the expected log-likelihood function
\begin{align}
\label{Q_single}
Q(x | x_k) = \sum_{m=0}^{N_d-1} \sum_{l=0}^{2L-1} p(l | y_m, x_k) \log p(y_m, l | x)
\end{align}
given the current signal estimate $x_k$, where
\begin{equation}
\label{prob_single}
p(y_m | l, x) \propto \prod_{i=0}^{L-1} \exp \bigg( \hspace{-3pt} -\frac{(y_m[i] - (C R_{l} Z x)[i])^2}{2\sigma^2} \bigg),
\end{equation}
with the normalization $\sum_{l=0}^{2L-1} p(y_m | l, x) = 1$.
From Bayes' rule, we have
\begin{align}
\label{bayes}
p(l | y_m, x_k) = \frac{p(y_m | l, x_k) p(l | x_k)}{\sum_{l=0}^{2L-1} p(y_m | l, x_k) p(l | x_k)},
\end{align}
which is the normalized likelihood function $p(y_m | l, x_k)$, weighted by the prior distribution $p(l | x_k)$.
In general, the prior distribution $p(l | x_k)$ is independent of the model $x_k$ and can be estimated simultaneously with the signal.

Denoting the prior distribution $p(l)$ by $\alpha[l]$, we rewrite \eqref{Q_single} as (up to an irrelevant constant)
\begin{align}
Q(x, \alpha | x_k, \alpha_k) &= \sum_{m=0}^{N_d-1} \sum_{l=0}^{2L-1} p(l | y_m, x_k) \nonumber \\
		&\hspace{3em} \times \bigg( \log p(y_m | l, x) + \log \alpha[l] \bigg). \nonumber
\end{align}
We note that the dependence of $Q(x, \alpha | x_k, \alpha_k)$ on the current prior estimates $\alpha_k$ lies in $p(l | y_m, x_k)$ through \eqref{bayes}.
The M-step updates the signal estimate and the priors by maximizing $Q(x, \alpha | x_k, \alpha_k)$ under the constraint that the priors lie on the simplex $\Delta_{2L}$:
\begin{equation}
\label{max_single}
x_{k+1}, \alpha_{k+1} = \arg\max_{x, \alpha} Q(x, \alpha | x_k, \alpha_k)~\text{s.t.}~\alpha \in \Delta_{2L}.
\end{equation}
As shown in Appendix~\ref{derv_update_single}, we obtain the update rules
\begin{align}
\label{update_single1}
\hspace{-8pt} x_{k+1}[j] = \frac{\sum_{m=0}^{N_d-1} \sum_{l=j+1}^{j+L} p(l | y_m, x_k) y_m[j+L-l]}{\sum_{m=0}^{N_d-1} \sum_{l=j+1}^{j+L} p(l | y_m, x_k)},
\end{align}
where $0 \leq j < L$, and
\begin{align}
\label{update_single2}
\alpha_{k+1}[l] = \frac{1}{N_d} \sum_{m=0}^{N_d-1} p(l | y_m, x_k),
\end{align}
where $0 \leq l < 2L$.
We repeat the iterations of the EM algorithm until the estimates stop improving, as judged within some tolerance.

\subsection{Arbitrary spacing distribution}
We extend the approximate EM approach to address arbitrary spacing distribution of signal occurrences by reformulating the probability model: each segment $y_m$ can now contain up to two signal occurrences.
In this case, the two signals can appear in $L(L-1)/2$ different combinations in a segment, which is explicitly modeled by
\begin{align}
y_m = C R_{l^m_1} Z x + C R_{l^m_2} Z x + \varepsilon_m,\hspace{5pt}\varepsilon_m \sim \mathcal{N}(0, \sigma^2 I_L), \nonumber
\end{align}
where $L < l^m_1 < 2L$ and $0 < l^m_2 \leq l^m_1 - L$.
Given the signal estimate $x_k$ and the shifts $l^m_1=l_1, l^m_2=l_2$, the likelihood function $p(y_m | l_1, l_2, x_k)$ can be written as
\begin{align}
\label{prob_double}
&p(y_m | l_1, l_2, x_k) \propto \prod_{i=0}^{L-1} \exp \bigg( -\frac{1}{2\sigma^2} \nonumber \\
	&\hspace{2em} \times \bigg[ y_m[i] - (C R_{l_1} Z x_k)[i] - (C R_{l_2} Z x_k)[i] \bigg]^2 \bigg),
\end{align}
and we have the normalization condition
\begin{align}
\sum_{l=0}^{2L-1} p(y_m | l, x_k) + \sum_{l_1=L+1}^{2L-1} \sum_{l_2=1}^{l_1-L} p(y_m | l_1, l_2, x_k) = 1. \nonumber
\end{align}
Incorporating the terms with two signal occurrences, the E-step constructs the expected log-likelihood function as
\begin{align}
\label{Q_double}
&Q(x, \alpha | x_k, \alpha_k) \nonumber \\
&= \sum_{m=0}^{N_d-1} \bigg[ \sum_{l=0}^{2L-1} p(l | y_m, x_k) \bigg( \log p(y_m | l, x) + \log\alpha[l] \bigg) \nonumber \\
	&\hspace{6pt} + \sum_{l_1=L+1}^{2L-1} \sum_{l_2=1}^{l_1-L} p(l_1, l_2 | y_m, x_k) \nonumber \\
	&\hspace{5em} \times \bigg( \log p(y_m | l_1, l_2, x) + \log \alpha[l_1, l_2] \bigg) \bigg],
\end{align}
where the prior $p(l_1, l_2)$ is denoted by $\alpha[l_1, l_2]$.

Under the assumption that the signal occurrences in any subset of $y$ follow the same spatial distribution, we can parametrize the priors with the pair separation function $\xi$ (see Definition~\ref{def:psf}).
Recall that $(M-1)\xi[l]$ is the number of pairs of consecutive signal occurrences whose starting positions are separated by exactly $l$ positions.
The priors $\alpha[l_1, l_2]$ can be related to the probability that two signal occurrences appear in the combination specified by $(l_1, l_2)$ in a segment of length $L$ selected from the measurement $y$, which is estimated by
\begin{align}
\label{prior1}
\alpha[l_1, l_2] = \frac{(M-1)\xi[l_1 - l_2]}{N-L+1} \approx \frac{M}{N}\xi[l_1 - l_2],
\end{align}
where $L < l_1 < 2L$ and $0 < l_2 \leq l_1-L$.
Here, $(M-1)\xi[l_1 - l_2]$ is the number of segments that realize the configuration of signal occurrences specified by $(l_1, l_2)$, and $N-L+1$ indicates the total number of segment choices.
The priors $\alpha[l]$ can similarly be related to the probability that a signal occurs in the way specified by $l$ in a segment of length $L$:
\begin{align}
\label{prior2}
\alpha[l] &= \alpha[2L-l] \nonumber \\
	&= \frac{(M-1)\xi[2L-l] + (M-1)\xi[2L-l+1] + \cdots}{N-L+1} \nonumber \\
	&\approx \frac{M}{N}\sum_{j=2L-l}^{\infty} \xi[j],
\end{align}
where $0 < l \leq L$.
An interesting observation is that the number of signal occurrences $M$ can be estimated by
\begin{align}
\alpha[L] \approx \frac{M}{N} \sum_{j = L}^{\infty} \xi[j] = \frac{M}{N}, \nonumber
\end{align}
since the signal occurrences are required not to overlap.
The value of the prior $\alpha[0]$ is determined by the normalization
\begin{align}
\label{asd_normalization}
\sum_{l=0}^{2L-1} \alpha[l] + \sum_{l_1=L+1}^{2L-1} \sum_{l_2=1}^{l_1-L} \alpha[l_1, l_2] = 1.
\end{align}
From \eqref{prior1}, \eqref{prior2} and \eqref{asd_normalization}, we see that the priors are uniquely specified by the positive parameters $\alpha[0]$, $\alpha[1]$, $\rho_1[0]$, $\rho_1[1]$, $\dots, \rho_1[L-2]$, with $\rho_1$ defined in \eqref{rho1}.
Therefore, the normalization \eqref{asd_normalization} can be rewritten as
\begin{align}
\label{constraint}
\alpha[0] + (2L-1)\alpha[1] + \sum_{i=0}^{L-2} \frac{i+L}{L} \rho_1[i] = 1.
\end{align}
In the special case of well-separated signals, where $\xi[i] = 0$ for $L \leq i \leq 2L-2$, we have $\alpha[1] = \alpha[2] = \cdots = \alpha[2L-1]$ and the normalization $\alpha[0] + (2L-1)\alpha[1] = 1$.

In the M-step, we update the signal estimate and the parameters $\alpha[0], \alpha[1], \rho_1$ by maximizing $Q(x, \alpha | x_k, \alpha_k)$ under the constraint that \eqref{constraint} is satisfied:
\begin{align}
\label{max_double}
&\hspace{-3pt}x_{k+1}, \alpha_{k+1}[0], \alpha_{k+1}[1], \{\rho_1\}_{k+1} \nonumber \\
&\hspace{-6pt}= \arg \max_{x, \alpha[0], \alpha[1], \rho_1} Q(x, \alpha | x_k, \alpha_k)~\text{s.t.}~\eqref{constraint}~\text{is satisfied}.
\end{align}
As shown in \eqref{Q_double}, $Q(x, \alpha | x_k, \alpha_k)$ is additively separable for $x$ and $\alpha$, so the constrained maximization (\ref{max_double}) can be reached by maximizing $Q(x, \alpha | x_k, \alpha_k)$ with respect to $x$ and the parameters $\alpha[0], \alpha[1], \rho_1$ separately.
Maximizing $Q(x, \alpha | x_k, \alpha_k)$ with respect to $x$ yields the update rule (see Appendix~\ref{derv_update_double} for derivation)
\begingroup
\allowdisplaybreaks
\begin{align}
\label{update_double}
&x_{k+1}[j] \nonumber \\
&= \bigg[ \sum_{m=0}^{N_d-1} \bigg( \sum_{l=j+1}^{j+L} p(l | y_m, x_k) y_m[j+L-l] \nonumber \\
	&\hspace{6pt} + \sum_{l_1=L+1}^{j+L} \sum_{l_2=1}^{l_1-L} p(l_1, l_2 | y_m, x_k) y_m[j+L-l_1] \nonumber \\
	&\hspace{6pt} + \sum_{l_1=L+j+1}^{2L-1} \sum_{l_2=j+1}^{l_1-L} p(l_1, l_2 | y_m, x_k) y_m[j+L-l_2] \bigg) \bigg] \nonumber \\
	&\times \bigg[ \sum_{m=0}^{N_d-1} \bigg( \sum_{l=j+1}^{j+L} p(l | y_m, x_k) + \sum_{l_1=L+1}^{j+L} \sum_{l_2=1}^{l_1-L} p(l_1, l_2 | y_m, x_k) \nonumber \\
	&\hspace{6pt} + \sum_{l_1=L+j+1}^{2L-1} \sum_{l_2=j+1}^{l_1-L} p(l_1, l_2 | y_m, x_k) \bigg) \bigg]^{-1},
\end{align}
\endgroup
where $0 \leq j < L$.
To update the parameters $\alpha[0], \alpha[1], \rho_1$, we note that the function $Q(x, \alpha | x_k, \alpha_k)$ is concave with respect to these parameters and the constraint~\eqref{constraint} forms a compact convex set.
Therefore, the constrained maximization is achieved using the Frank-Wolfe algorithm~\cite{Frank56}.

\subsection{Frequency marching}
Because the iterates of the EM algorithm are not guaranteed to converge to the maximum likelihood, we develop a frequency marching scheme to help the iterates converge to the global maximum.
Recall that frequency marching converts the original optimization problem into a series of coarse-grained problems with gradually increasing resolution.
The coarse-grained version of the EM algorithm is characterized by the spatial resolution
\begin{align}
\Delta x = \max(1, L/2n_{\max}), \nonumber
\end{align}
where $n_{\max} = 1, 2, \dots, \lfloor (L+1)/2 \rfloor$.
This spatial resolution models the signal shifts in the coarse-grained problem, with the corresponding expected likelihood function given by
\begin{align}
& Q^{(n_{\max})}(x, \alpha^{(n_{\max})} | x_k, \alpha^{(n_{\max})}_k) \nonumber \\
& = \sum_{m=0}^{N_d-1} \bigg[ \sum_{l=0}^{2L'-1} p(\lfloor l\Delta_x \rceil | y_m, x_k) \bigg( \log p(y_m | \lfloor l\Delta_x \rceil, x) \nonumber \\
	&+ \log\alpha^{(n_{\max})}[l] \bigg) + \hspace{-0.25em} \sum_{l_1=L'+1}^{2L'-1} \sum_{l_2=1}^{l_1-L'} p(\lfloor l_1\Delta_x \rceil, \lfloor l_2\Delta_x \rceil | y_m, x_k) \nonumber \\
	&\times \bigg( \log p(y_m | \lfloor l_1\Delta_x \rceil, \lfloor l_2\Delta_x \rceil, x) + \log \alpha^{(n_{\max})}[l_1, l_2] \bigg) \bigg], \nonumber
\end{align}
where $L' = L/\Delta_x$.
This likelihood function is equal to the original one shown in (\ref{Q_double}) restricted to the rounded shifts, so the signal can be updated with (\ref{update_double}) by ignoring the terms with irrelevant shifts.

Mimicking \eqref{prior1} and \eqref{prior2}, we construct the expressions for the priors in the coarse-grained problem as
\begin{align}
\alpha^{(n_{\max})}[l_1, l_2] &\approx \frac{M}{N} \xi^{(n_{\max})}[l_1 - l_2] \nonumber \\
\alpha^{(n_{\max})}[l] = \alpha^{(n_{\max})}[2L'-l] &\approx \frac{M}{N} \hspace{-0.3em} \sum_{j = 2L'-l}^{\infty} \xi^{(n_{\max})}[j], \nonumber
\end{align}
where $L' < l_1 < 2L'$, $0 < l_2 \leq l_1-L'$, $0 < l \leq L'$, and $\xi^{(n_{\max})}$ is defined in \eqref{coarse_psf}.
Since the priors are uniquely specified by the positive parameters $\alpha^{(n_{\max})}[0],$ $\alpha^{(n_{\max})}[1],$ $\rho^{(n_{\max})}_1[0],$ $\rho^{(n_{\max})}_1[1],$ $\dots, \rho^{(n_{\max})}_1[L'-2],$ with $\rho_1^{(n_{\max})}$ defined in \eqref{rho1_coarse}, we can update the priors by maximizing $Q^{(n_{\max})}(x, \alpha^{(n_{\max})} | x_k, \alpha^{(n_{\max})}_k)$ with respect to these parameters under the constraint that the parameters lie on the simplex defined by
\begin{align}
\alpha^{(n_{\max})}[0] + (2L'-1)\alpha^{(n_{\max})}[1] + \sum_{i=0}^{L'-2} \frac{i+L'}{L'} \rho^{(n_{\max})}_1[i] = 1. \nonumber
\end{align}
Incrementing from $n_{\max} = 1$ to $n_{\max} = \lfloor (L+1)/2 \rfloor$, the estimated signal and priors in each stage of frequency marching are used to initialize the optimization problem in the next stage.

\section{Numerical Experiments}
\label{section:numerical}
\begin{figure}[t]
\includegraphics[scale=0.65]{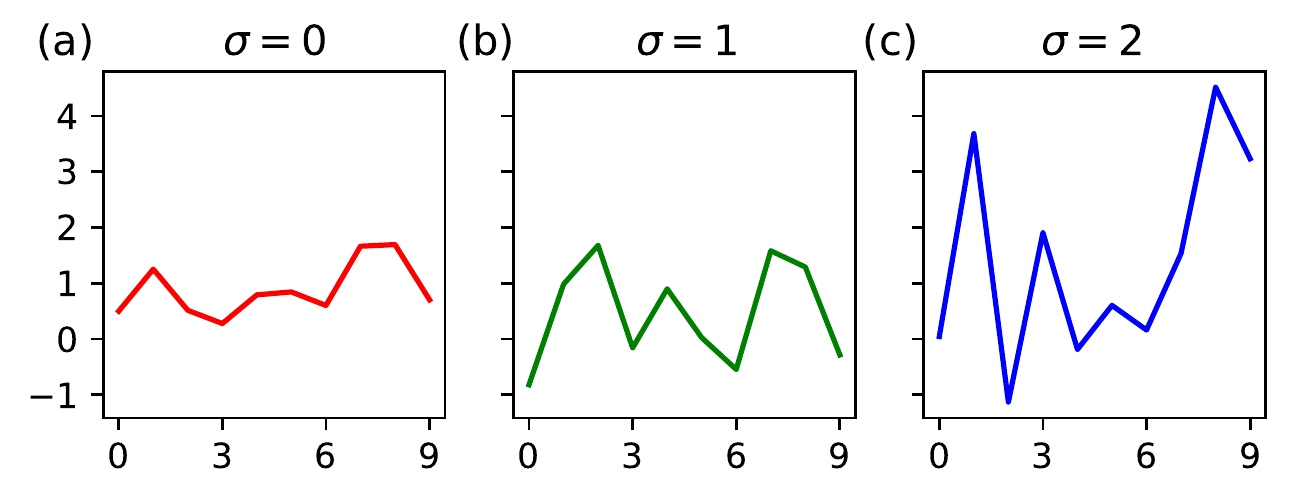}
\caption{(a) The signal used to generate our synthetic data, with the $\ell_2$ norm scaled to $\sqrt{L}$.
(b), (c) Signals corrupted by additive white Gaussian noise with zero mean and standard deviations $\sigma = 1$ and $\sigma = 2$.}
\label{fig:signal}
\end{figure}

\begin{figure}[h]
\centering
\includegraphics[scale=0.65, trim=0cm 0.25cm 0cm 0.25cm, clip=true]{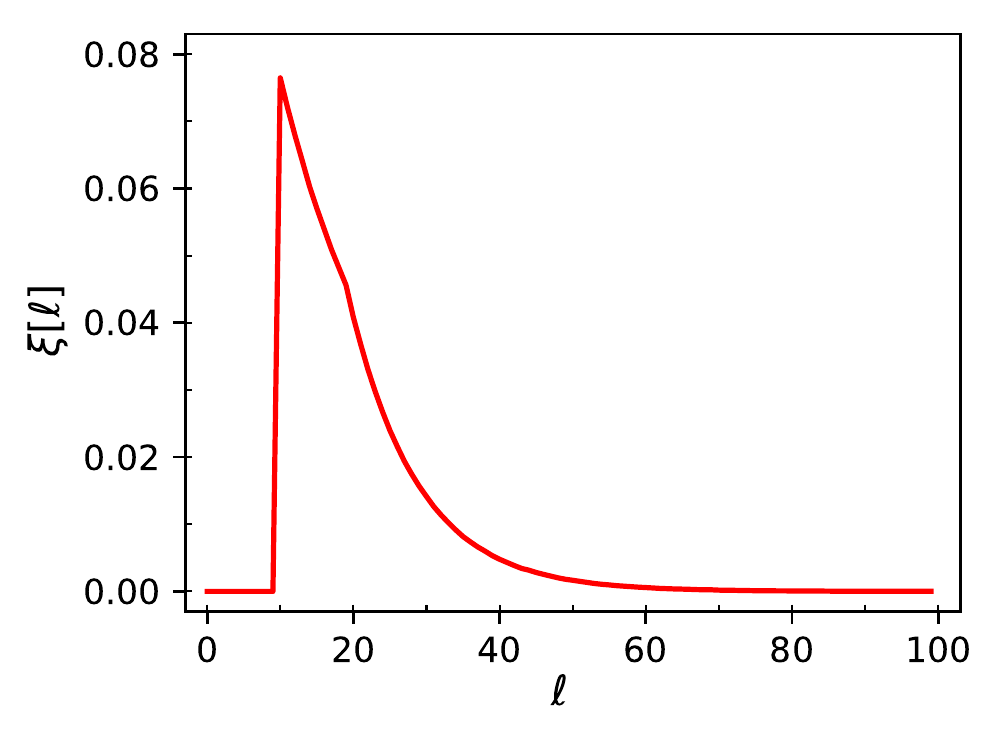}
\caption{Pair separation function to test arbitrary signal spacing distributions.}
\label{fig:psf}
\end{figure}

This section describes the construction of our synthetic data and the performance of the two proposed methods.\footnote{The code for all experiments is publicly available at \url{https://github.com/tl578/multi-target-detection}.}
A measurement is generated by first sampling $M$ integers from $[0, N-L]$, with the constraint that any two samples differ by at least $L+W$ entries, for some integer $W$.
We use $W = L-1$ for the case of well-separated signals, while $W = 0$ is chosen to test our methods for arbitrary spacing distribution of signal occurrences.
The sampling is done by generating the random integers uniformly one by one, rejecting any integer that would violate the constraint with respect to previously accepted samples.
The $M$ integers indicate the starting positions of the signal occurrences, which are recorded as the non-zero entries of the binary sequence $s$.
The noisy measurement $y$ is given by the linear convolution of $s$ with the signal $x$, with each entry of $y$ further corrupted by additive white Gaussian noise of zero mean and variance $\sigma^2$.
Our synthetic data all have length $N = 10^6$, and are constructed with the signal shown in Figure~\ref{fig:signal}(a), which has length $L = 10$.
Also shown in Figures~\ref{fig:signal}(b) and~\ref{fig:signal}(c) are examples of the corrupted signals by different levels of noise.
The number of signal occurrences $M$ is adjusted so that the signal density $\rho_0 = ML/N$ is equal to 0.3 and 0.5 for the cases of well-separated signals and arbitrary spacing distribution respectively. 
Figure~\ref{fig:psf} shows the resulting pair separation function for our test case of arbitrary spacing distribution.

For both cases of signal spacing distribution, we reconstruct the signal from synthetic data using the two proposed methods with the frequency marching scheme over a wide range of $\sigma$.
When the frequency marching scheme is not implemented, we observe that the algorithms usually suffer from large errors due to the non-convexity of the problems.
A total of 20 instances of synthetic data are generated for each value of $\sigma$, and each instance is solved by the two methods using the frequency marching schemes.
For our autocorrelation analysis methods, we run the optimization with 10 different random initializations and retain the one that minimizes the cost function, as defined in \eqref{cost_ws} or \eqref{cost_asd}.
For the EM approach, we also use 10 random initializations and retain the solution with the maximum data likelihood.
Given the signal estimate~$x_k$, we approximate the data likelihood for the cases of well-separated signals and arbitrary spacing distribution by
\begin{align}
p(y | x_k) \approx \prod_{m=0}^{N_d-1} \sum_{l=0}^{2L-1} p(y_m | l, x_k) \alpha[l] \nonumber
\end{align}
and
\begin{align}
p(y | x_k) &\approx \prod_{m=0}^{N_d-1} \bigg( \sum_{l=0}^{2L-1} p(y_m | l, x_k) \alpha[l] \nonumber \\
	&\hspace{2em} + \sum_{l_1=L+1}^{2L-1} \sum_{l_2=1}^{l_1-L} p(y_m | l_1, l_2, x_k) \alpha[l_1, l_2] \bigg) \nonumber
\end{align}
respectively.
This is the same approximation we use to formulate the EM algorithm.

\begin{figure}[h]
\includegraphics[scale=0.56, trim=0cm 0.3cm 0cm 0.15cm, clip=true]{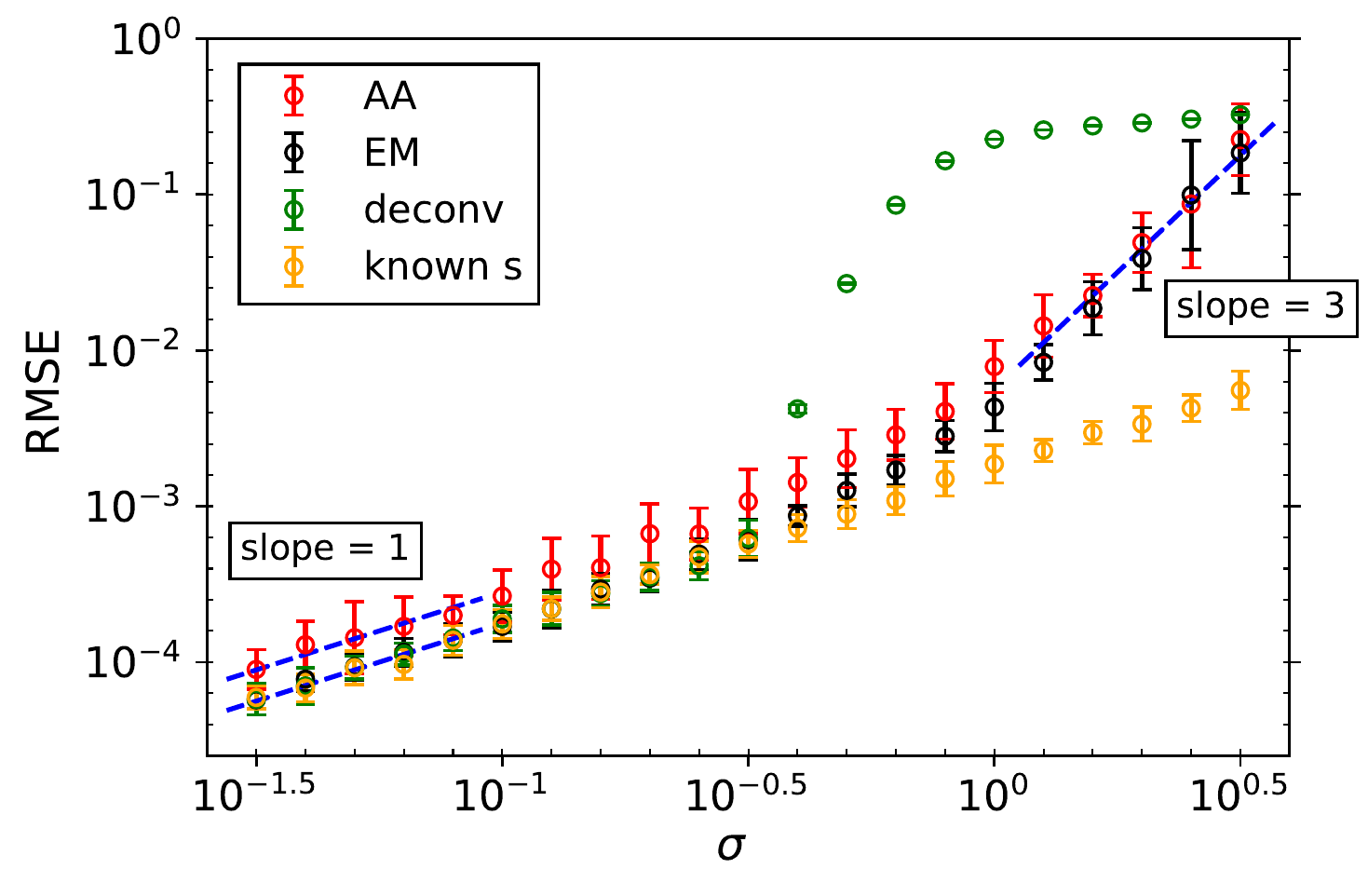}
\centering
\caption{The RMSEs of the reconstructed signal from data generated with well-separated occurrences. We show performance of our autocorrelation analysis (AA) and of expectation-maximization (EM). Also shown are the RMSEs of the signal estimated by the oracle-based deconvolution algorithm `deconv' described in Section~\ref{section:numerical} and the RMSEs of the estimated signal when the binary sequence $s$ is known.}
\label{fig:ws}
\end{figure}

\begin{figure}[h]
\includegraphics[scale=0.65]{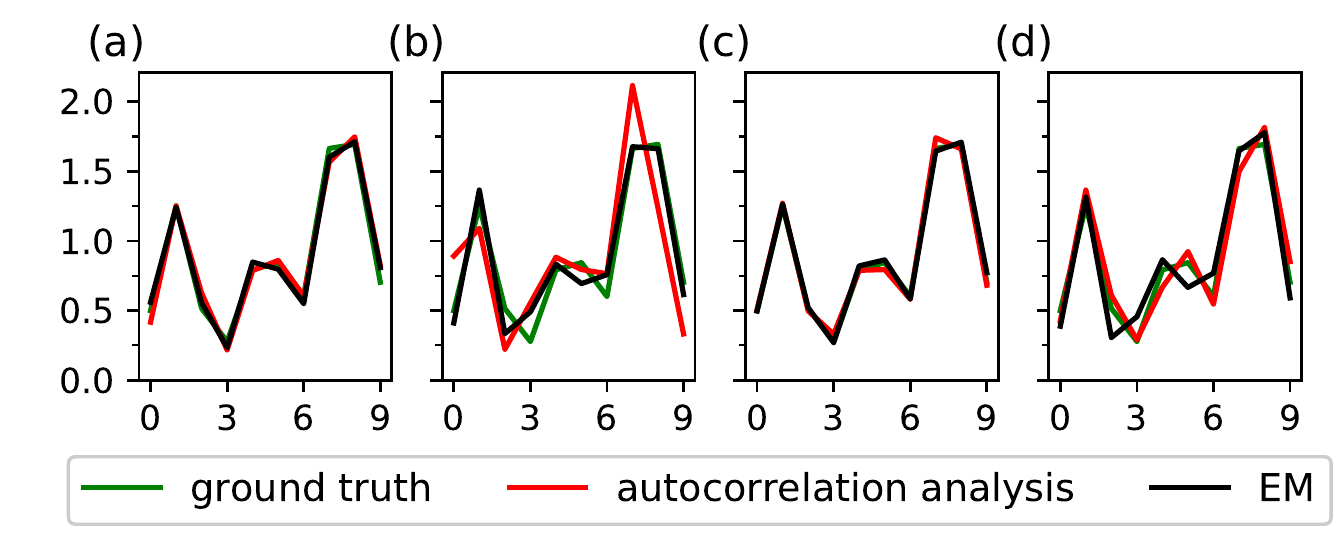}
\centering
\caption{(a), (b) An instance of the reconstructed signals from data generated with well-separated occurrences for $\sigma = 10^{0.3}$ and $\sigma = 10^{0.5}$ respectively. (c), (d) An instance of the reconstructed signals from data generated with an arbitrary spacing distribution for $\sigma = 10^{0.3}$ and $\sigma = 10^{0.5}$ respectively.}
\label{fig:recon}
\end{figure}

The RMSEs of the reconstructions for the case of well-separated signals are shown in Figure~\ref{fig:ws}.
As predicted in Section~\ref{section:AA_ws}, the RMSE for autocorrelation analysis scales in proportion to $\sigma$ at high SNR, and to $\sigma^3$ at low SNR.
Interestingly, the EM algorithm exhibits the same behavior, which empirically shows that the two approaches share the same scaling of sample complexity in the two extremes of noise levels.
Figure~\ref{fig:recon}(a) and~\ref{fig:recon}(b) show an instance of the signals reconstructed from data generated with $\sigma = 10^{0.3}$ and $\sigma = 10^{0.5}$ respectively.
The same scaling is not observed for the reconstructed signal density $\rho_0$ (not shown), although the relative errors are generally well below a few percents even in the noisiest cases.

\begin{figure}[h]
\includegraphics[scale=0.56, trim=0cm 0.3cm 0cm 0.15cm, clip=true]{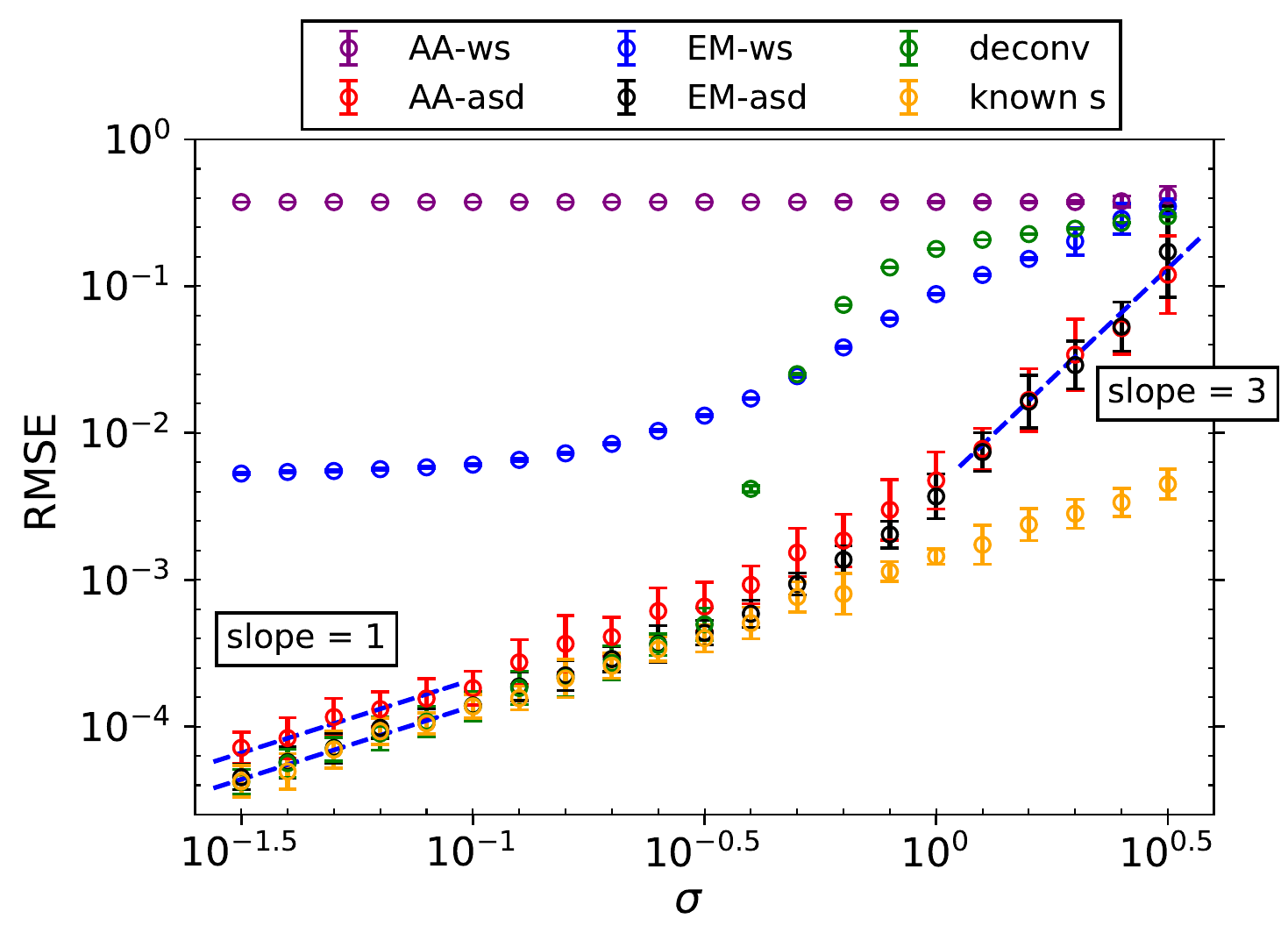}
\centering
\caption{The RMSEs of the reconstructed signal from data generated with an arbitrary spacing distribution. We show performance of our autocorrelation analysis (AA) and of expectation-maximization (EM), both under the (incorrect) assumptions of well-separated signals (ws) and the (correct) arbitrary spacing distribution model (asd).
Also shown are the RMSEs of the signal estimated by the oracle-based deconvolution algorithm `deconv' described in Section~\ref{section:numerical} and the RMSEs of the estimated signal when the binary sequence $s$ is known.}
\label{fig:asd}
\end{figure}

Figure~\ref{fig:asd} shows the RMSEs of the reconstructed signal for an arbitrary spacing distribution displayed in Figure~\ref{fig:psf}.
As a comparison, we also run the estimation algorithms under the (incorrect) assumption of well-separated signals on the same datasets.
We first see that autocorrelation analysis assuming well-separated signals obtains poor reconstruction at all noise levels.
Although the EM approach that assumes well-separated signals produces more accurate estimates, the nearly constant RMSEs at high SNR indicate the systematic error due to model misspecification.
By contrast, the algorithms that assume an arbitrary spacing distribution achieve much better reconstruction, and the resulting RMSEs have the same scaling behaviors at the two extremes of SNR as shown in Figure~\ref{fig:ws}.
An instance of the signals reconstructed from data generated with $\sigma = 10^{0.3}$ and $\sigma = 10^{0.5}$ are shown in Figure~\ref{fig:recon}(c) and~\ref{fig:recon}(d) respectively.

One of the premises of this paper is that, at high noise levels, detection-based methods are destined to fail because detection cannot be done reliably. To support this theoretical argument further, in Figures~\ref{fig:ws} and~\ref{fig:asd} we display the performance of the following oracle, named `deconv'. This method is given a strictly simpler task to solve: it must estimate $x$ given the observation $y$, the noise level $\sigma$ and the number of signal occurrences $M$, as well as the $\ell_2$-distance between the ground truth $x$ and every single window of length $L$ in the observation~$y$. Precisely, the oracle has access to the vector $z \in \mathbb{R}^{N-L+1}$ defined for $i = 0, \ldots, N-L$ by
\begin{align*}
	z[i] & = \sum_{l = 0}^{L-1} \left( x[l] - y[i+l] \right)^2.
\end{align*}
The vector $z$ provides this estimator the advantage of finding the likely locations of signal occurrences.
If the norm of $x$ is known, this oracle advantage is equivalent to knowledge of the cross-correlation between $x$ and $y$. Then, the oracle-based estimator `deconv' proceeds as follows: it selects the index $i$ corresponding to the lowest value in $z$, excludes other indices too close to $i$ as prescribed by the signal separation model under consideration, and repeats this procedure $M$ times to produce an estimator of $s$: the $M$ starting positions of signal occurrences in $y$. The signal $x$ is then estimated by averaging the $M$ selected segments of $y$, effectively deconvolving $y$ by the estimator of $s$. To our point: despite the unfair oracle advantage, this estimator is unable to estimate $x$ to a competitive accuracy when the noise level is large.

Also plotted in Figures~\ref{fig:ws} and~\ref{fig:asd} are the RMSEs of the estimated signal when the binary sequence $s$ is known. 
In this case, the RMSEs scale as $\sigma$ and serve as the lower bounds of any algorithm. 
We can see that, at the low noise levels, our approximate EM algorithm performs nearly as well as the case when $s$ is known.

\begin{figure}[h]
\includegraphics[scale=0.56, trim=0cm 0.3cm 0cm 0.15cm, clip=true]{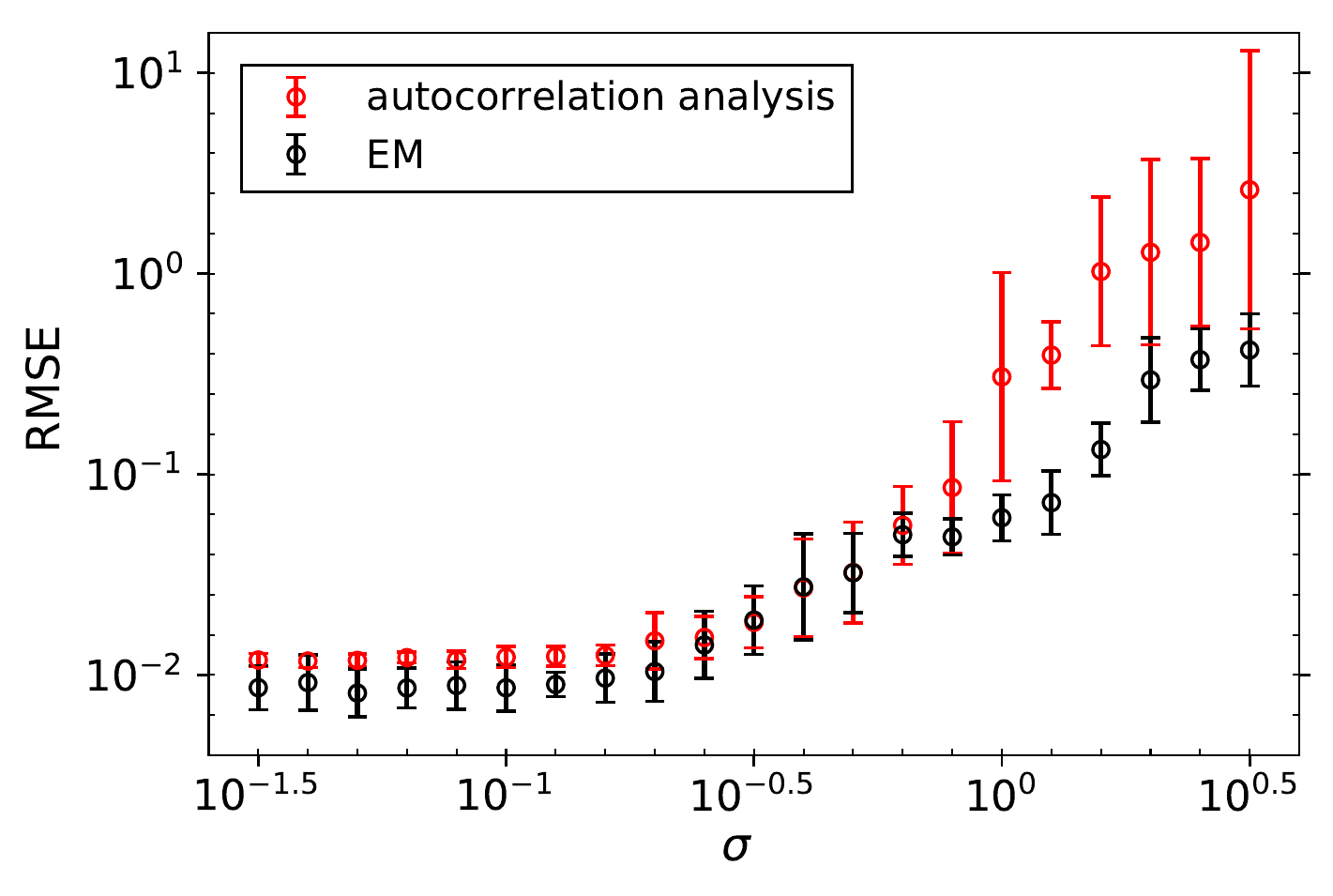}
\centering
\caption{The (relative) RMSEs of the reconstructed values of $\rho_1$ as defined by~\eqref{rho1}, for the case of arbitrary spacing distribution.}
\label{fig:rho1}
\end{figure}

Figure~\ref{fig:rho1} shows the RMSEs of the reconstructed values of $\rho_1$ for the experiments ``AA-asd'' and ``EM-asd'' shown in Figure~\ref{fig:asd}.
Although our methods are not able to reconstruct $\rho_1$ to high precision, the results shown in Figure~\ref{fig:asd} indicate the necessity to include the pair separation function in the model to achieve good signal reconstruction.

\begin{figure}[h]
\includegraphics[scale=0.56, trim=0cm 0.3cm 0cm 0.3cm, clip=true]{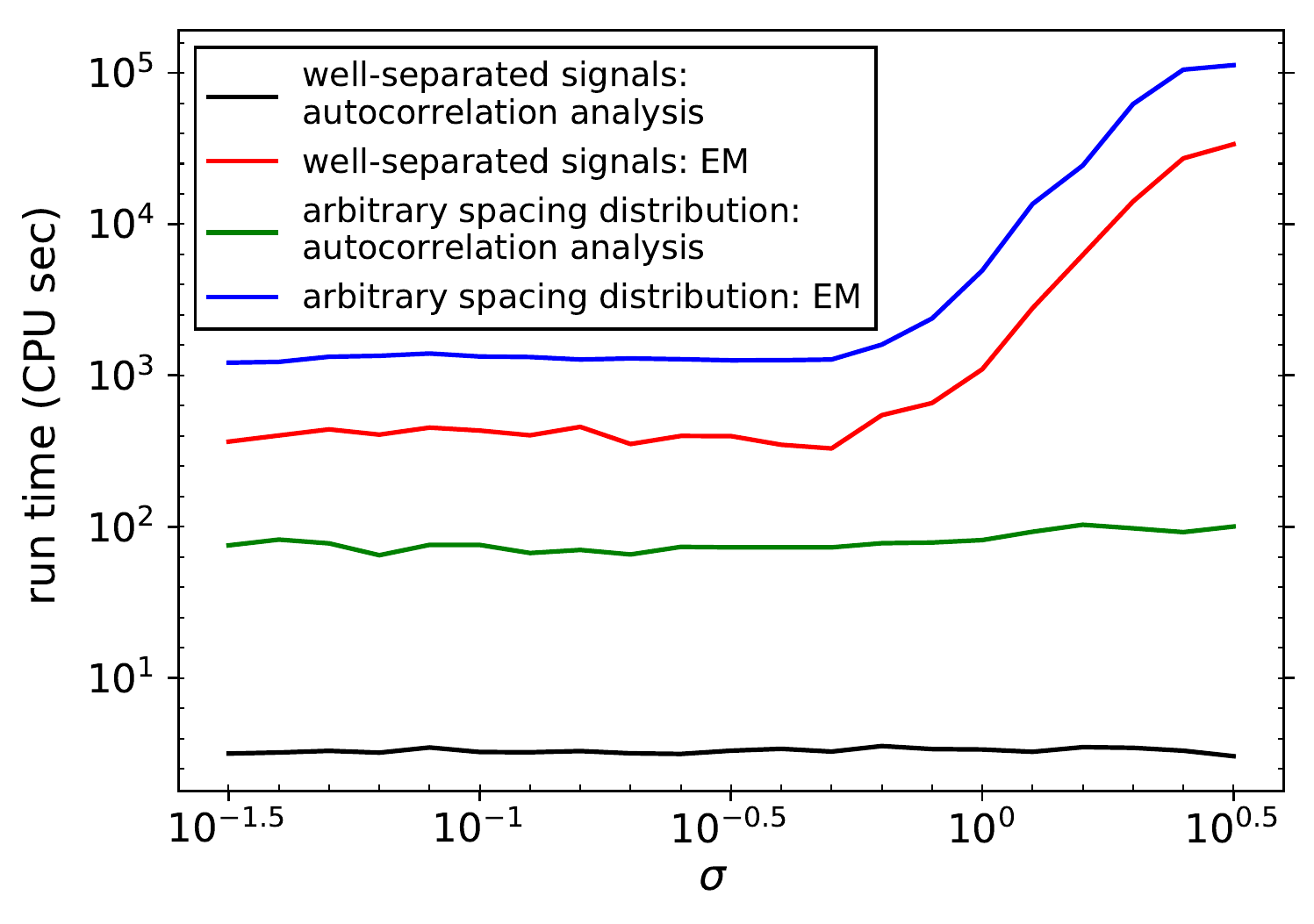}
\centering
\caption{The average run time to solve an instance of synthetic data for the two methods in both cases of signal spacing distribution.}
\label{fig:run_time}
\end{figure}

Figure~\ref{fig:run_time} shows the average computation time for the main methods in Figures~\ref{fig:ws} and~\ref{fig:asd}.
EM is slower because it cross-correlates all the observed segments with the signal estimate in each iteration.
This issue becomes especially prominent in the case of arbitrary spacing distribution. On the other hand, autocorrelation analysis requires shorter computation time by summarizing the data as autocorrelation statistics with one pass over the data.
We note that the distinct computational speeds for autocorrelation analysis and the EM algorithm is also observed in the related problem of multi-reference alignment~\cite{Bendory18a,Boumal18}.
The difference in run time and the similar reconstruction quality of the two methods at low SNR make autocorrelation analysis the preferred approach for large datasets.

\section{Discussion}
\label{section:discussion}
In this study, we have presented two approaches---autocorrelation analysis and an approximate EM algorithm---to tackle the MTD problem without the need to detect the positions of the underlying signal occurrences.
By incorporating the notion of pair separation function, we generalize the solution of the MTD problem from the special case of well-separated signals to allow an arbitrary spacing distribution of signal occurrence.
It is empirically shown that the two methods have the same scaling of sample complexity in the two extremes of noise levels, in particular, SNR${}^{-3}$ at high noise level.
Since the optimizations in both methods are non-convex, computing schemes based on frequency marching are designed to help the iterates converge to the global optimum.

Our study of the MTD model is primarily motivated by the goal to reconstruct the structures of small biomolecules using cryo-EM~\cite{Frank06,Bendory19b}.
In a cryo-EM experiment, individual copies of the target biomolecule are dispersed at unknown 2D locations and 3D orientations in a thin layer of vitreous ice, from which 2D tomographic projection images are produced by an electron microscope.
To minimize the irreversible structural damage, it is necessary to keep the electron dose low.
As a result, the projection images are considerably noisy, and high-resolution structure estimation requires averaging over a large number of noisy projections.
In particular, as the size of the molecule gets smaller, the SNR of the data drops correspondingly.

Currently, the analysis workflow of cryo-EM data is roughly divided into two steps:
The first step, known as particle picking, detects the locations of the biomolecules in the noisy projection images.
The 3D structure of the biomolecule is reconstructed in the second step from the unoriented particle projections.
When the sizes of biomolecules get smaller, however, reliable detection of their positions becomes challenging, which in turn hampers successful particle picking~\cite{Aguerrebere16,Bendory18b}.
It is estimated from first principles that, in order to obtain a~3~\AA~resolution reconstruction, the particle size should be at least 45~\AA~so that the particle occurrences can be accurately detected~\cite{Henderson95}.
The results of \cite{Bendory18b,Bendory19} and this work suggest that it is possible to bypass the need to detect particle positions but still reconstruct the structure at high resolution.

A key feature of cryo-EM that the MTD model fails to capture is that the ``signals'' are actually 2D projections of the underlying biomolecules at random 3D orientations, and the orientation distribution is usually non-uniform and unknown.
This distinction makes the direct application of our approximate EM algorithm prohibitive, because each observed window now needs to cross-correlate with model projections with all possible 2D in-plane translations and 3D orientations, let alone the case where multiple copies of biomolecules may appear in a window.
Further approximations such as representing the data and model projections in low-dimensional subspaces~\cite{Dvornek15,Landa18} seem necessary to push this approach forward.
We note that the use of frequency marching in the MTD problem is also related to this idea: the signal is mapped to low-dimensional subspaces with growing dimensions specified by $n_{\max}$.

On the other hand, the autocorrelations calculated from the noisy measurements in cryo-EM involve averages over both the 3D orientations and 2D in-plane translations.
It was pointed out that the 3D structure reconstruction problem by fitting the first three autocorrelations of a biomolecule might be ill-conditioned~\cite{Bendory18b}, so information from higher-order autocorrelations seems required.
We also expect the cross-terms between 2D projections of different copies to further complicate the reconstruction problem.
The resolution of these technical challenges may help extend the use of cryo-EM to smaller biomolecules that are currently believed insurmountable, and is the subject of our ongoing studies.

\section*{Acknowledgment}
We would like to thank Ayelet Heimowitz, Joe Kileel, Eitan Levin and Amit Moscovich for stimulating discussions and the anonymous reviewers for their valuable comments.

\appendices

\section{Derivations of Relations \eqref{a1_ws}--\eqref{a3_ws} and \eqref{a1_asd}--\eqref{a3_asd}}
\label{aa_relations}
With the noisy measurement $y$ defined in \eqref{MTD}, we express its first order autocorrelation as
\begingroup
\allowdisplaybreaks
\begin{align}
a_y^1 &= \frac{1}{N} \sum_{i=0}^{N-1} (s \ast x)[i] + \frac{1}{N} \sum_{i=0}^{N-1} \varepsilon[i] \nonumber \\
	&= \frac{M}{N} \sum_{i=0}^{L-1} x[i] + \frac{1}{N} \sum_{i=0}^{N-1} \varepsilon[i] 
	= \rho_0 a_x^1 + \frac{1}{N} \sum_{i=0}^{N-1} \varepsilon[i]. \nonumber
\end{align}
\endgroup
Taking the expectation of $a_y^1$ with respect to the distribution of Gaussian noise, we obtain
\begin{align}
\mathop{\mathbb{E}_\varepsilon} \{a_y^1\} = \rho_0 a_x^1 \nonumber,
\end{align}
with variance $\mathcal{O}(\sigma^2/N)$.
We note that this expression applies to both the cases of well-separated signals and arbitrary spacing distribution.

The second order autocorrelation of $y$ with shifts $l = 0, 1, \dots, L-1$ can be written as
\begingroup
\allowdisplaybreaks
\begin{align}
\label{ay2}
a_y^2[l] &= \frac{1}{N} \sum_{i=0}^{N-1} y[i] y[i+l] \nonumber \\
&= \frac{1}{N} \sum_{i=0}^{N-1} \bigg( (s \ast x)[i] (s \ast x)[i+l] + (s \ast x)[i] \varepsilon[i+l] \nonumber \\
&+ \varepsilon[i] (s \ast x)[i+l] + \varepsilon[i] \varepsilon[i+l] \bigg) \nonumber \\
&= a^2_{s \ast x}[l] + \frac{1}{N} \sum_{i=0}^{N-1} \varepsilon[i] \varepsilon[i+l] \nonumber \\
&+ \frac{1}{N} \sum_{k=0}^{M-1} \sum_{i=0}^{L-1} x[i] (\varepsilon[s_k-l+i] + \varepsilon[s_k+l+i]),
\end{align}
\endgroup
where $a^2_{s \ast x}[l]$ denotes the second order auto\-correlation of the linear convolution $s \ast x$ with shifts $l$, and $s_k$ indicates the index of the $k^{\mathrm{th}}$ non-zero entry of the sequence $s$.
Taking the expectation of $a_y^2[l]$ with respect to the distribution of Gaussian noise, we have
\begin{align}
\mathop{\mathbb{E}_\varepsilon} \{a_y^2[l]\} = a^2_{s \ast x}[l] + \sigma^2 \delta[l]. \nonumber
\end{align}
The term that scales as $\mathcal{O}\big(\varepsilon^2\big)$ in \eqref{ay2} contributes to the variance by $\mathcal{O}\big(\sigma^4/N\big)$, while the terms that scale as $\mathcal{O}\big(\varepsilon\big)$ contribute to the variance by $\mathcal{O}\big(M\sigma^2/N^2\big)$.
Since we assume that $M$ grows with $N$ at a constant rate, the resulting variance scales as $\mathcal{O}\big((\sigma^2 + \sigma^4)/N \big)$.

For well-separated signals,
\begin{align}
a^2_{s \ast x}[l] = \frac{M}{N} \sum_{i=0}^{L-1} x[i] x[i+l] = \rho_0 a^2_x[l], \nonumber
\end{align}
and we obtain the relation in (\ref{a2_ws}).
For arbitrary spacing distribution, we need to include the correlation terms between consecutive signal occurrences:
\begin{align}
a^2_{s \ast x}[l] &= \rho_0 a^2_x[l] + \frac{M}{N} \sum_{j=L}^{L+l-1} \xi[j] \sum_{i=0}^{L-1} x[i] x[i+j-l] \nonumber \\
	&= \rho_0 a_x^2[l] + \rho_0 \sum_{j=L}^{L+l-1} \xi[j] a_x^2[j-l], \nonumber
\end{align}
which gives the relation in (\ref{a2_asd}).
The quantity $\xi[j]$ denotes the number of consecutive 1's that are separated by exactly $j$ entries in the sequence $s$, divided by $M-1$ so that $\sum_j \xi[j] = 1$.

As for the third order autocorrelation of $y$ with shifts $0 \leq l_1 \leq l_2 < L$, we have
\begingroup
\allowdisplaybreaks
\begin{align}
\label{ay3}
&a_y^3[l_1, l_2] = \frac{1}{N} \sum_{i=0}^{N-1} y[i] y[i+l_1] y[i+l_2] \nonumber \\
&= \frac{1}{N} \sum_{i=0}^{N-1} \bigg( (s \ast x)[i] (s \ast x)[i+l_1] (s \ast x)[i+l_2] \nonumber \\
&+ (s \ast x)[i] (s \ast x)[i+l_1] \varepsilon[i+l_2] + (s \ast x)[i] \varepsilon[i+l_1] \varepsilon[i+l_2] \nonumber \\
&+ \varepsilon[i] (s \ast x)[i+l_1] (s \ast x)[i+l_2] + \varepsilon[i] (s \ast x)[i+l_1] \varepsilon[i+l_2] \nonumber \\
&+ (s \ast x)[i] \varepsilon[i+l_1] (s \ast x)[i+l_2] + \varepsilon[i] \varepsilon[i+l_1] (s \ast x)[i+l_2] \nonumber \\
&+ \varepsilon[i] \varepsilon[i+l_1] \varepsilon[i+l_2] \bigg) \nonumber \\
&= a^3_{s \ast x}[l_1, l_2] + \frac{1}{N} \sum_{i=0}^{N-1} \varepsilon[i] \varepsilon[i+l_1] \varepsilon[i+l_2] \nonumber \\
&+ \frac{1}{N} \sum_{k=0}^{M-1} \sum_{i=0}^{L-1} \bigg( x[i] \big( x[i+l_1] + \varepsilon[s_k+l_1+i] \big) \varepsilon[s_k+l_2+i] \nonumber \\
&+ \varepsilon[s_k-l_1+i] x[i] \big( x[i+l_2-l_1] + \varepsilon[s_k+l_2-l_1+i] \big) \nonumber \\
&+ \big( x[i-l_2] + \varepsilon[s_k-l_2+i] \big) \varepsilon[s_k-l_2+l_1+i] x[i] \bigg), 
\end{align}
\endgroup
where $a^3_{s \ast x}[l_1, l_2]$ denotes the third order auto\-correlation of the linear convolution $s \ast x$ with shifts $(l_1, l_2)$, and $x$ is zero-padded for indices out of the range $[0, L-1]$.

Taking the expectation of $a_y^3[l_1, l_2]$ with respect to the distribution of Gaussian noise, we obtain
\begingroup
\allowdisplaybreaks
\begin{align}
&\mathop{\mathbb{E}_\varepsilon} \{a_y^3[l_1, l_2]\} \nonumber \\
&= a^3_{s \ast x}[l_1, l_2] + \frac{M}{N} \sigma^2 \big( \delta[l_1] + \delta[l_2] + \delta[l_1-l_2] \big) \sum_{i=0}^{L-1} x[i] \nonumber \\
&= a^3_{s \ast x}[l_1, l_2] + \rho_0 a^1_x \sigma^2 \big( \delta[l_1] + \delta[l_2] + \delta[l_1-l_2] \big). \nonumber
\end{align}
\endgroup
As for the variance, the term that scales as $\mathcal{O}\big( \varepsilon^3 \big)$ in \eqref{ay3} contributes to the variance by $\mathcal{O}\big( \sigma^6/N \big)$, while the terms that scale as $\mathcal{O}\big( \varepsilon \big)$ contribute to the variance by $\mathcal{O}\big( M\sigma^2/N^2 \big)$. Since we assume that $M$ grows with $N$ at a constant rate, the resulting variance scales as $\mathcal{O}\big((\sigma^2 + \sigma^6)/N \big)$.

For well-separated signals, we have
\begin{align}
a^3_{s \ast x}[l_1, l_2] = \rho_0 a^3_x[l_1, l_2], \nonumber
\end{align}
and we obtain the relation in (\ref{a3_ws}).
By including the correlations between consecutive signal occurrences, we expand $a^3_{s \ast x}[l_1, l_2]$ for the case of arbitrary spacing distribution as
\begingroup
\allowdisplaybreaks
\begin{align}
&a^3_{s \ast x}[l_1, l_2] = \rho_0 a^3_x[l_1, l_2] \nonumber \\
&+ \frac{M}{N} \sum_{j = L}^{L+l_2-l_1-1} \hspace{-0.75em} \xi[j] \sum_{i=0}^{L-1} x[i+j-l_2] x[i+j+l_1-l_2] x[i] \nonumber \\
&+ \frac{M}{N} \sum_{j = L}^{L+l_1-1} \xi[j] \sum_{i=0}^{L-1} x[i+j-l_1] x[i] x[i+l_2-l_1] \nonumber \\
&= \rho_0 a^3_x[l_1, l_2] + \rho_0 \hspace{-0.75em} \sum_{j = L}^{L+l_2-l_1-1} \hspace{-0.75em} \xi[j] a_x^3[j-l_2, j+l_1-l_2] \nonumber \\
&+ \rho_0 \sum_{j = L}^{L+l_1-1} \xi[j] a_x^3[l_2-l_1, j-l_1], \nonumber
\end{align}
\endgroup
which gives the relation in \eqref{a3_asd}.

\section{Derivation of \eqref{update_single1} and \eqref{update_single2}}
\label{derv_update_single}
The constrained maximization in \eqref{max_single} can be achieved with the unconstrained maximization of the Lagrangian
\begin{align}
\mathcal{L}(x, \alpha, \lambda) = Q(x, \alpha | x_k, \alpha_k) + \lambda \bigg(1 - \sum_{l=0}^{2L-1} \alpha[l] \bigg), \nonumber
\end{align}
where $\lambda$ denotes the Lagrange multiplier.
We note that the constraints in \eqref{max_single} involve the inequalities that the priors are non-negative. 
Such constrained maximization in general cannot be achieved by maximizing the Lagrangian, for the inequalities might be violated. 
As we will see later, however, these inequalities are automatically satisfied at the computed maximum (or local maximum) of the Lagrangian, which justifies this approach.

Since $Q(x, \alpha | x_k, \alpha_k)$ is additively separable for $x$ and $\alpha$, we maximize $\mathcal{L}(x, \alpha, \lambda)$ with respect to $x$ and $\alpha$ separately.
At the maximum of $\mathcal{L}(x, \alpha, \lambda)$, we have
\begin{align}
\label{xderv_L}
0 = \frac{\partial \mathcal{L}}{\partial x[j]} = \sum_{m=0}^{N_d-1} \sum_{l=0}^{2L-1} p(l | y_m, x_k) \frac{\partial \log p(y_m | l, x)}{\partial x[j]},
\end{align}
where $0 \leq j < L$. With $p(y_m | l, x)$ give by \eqref{prob_single}, we can write
\begingroup
\allowdisplaybreaks
\begin{align}
\label{xderv_single}
&\frac{\partial \log p(y_m | l, x)}{\partial x[j]} \nonumber \\
&= - \frac{\partial}{\partial x[j]} \sum_{i=0}^{L-1} \frac{( y_m[i] - (C R_l Z x)[i] )^2}{2\sigma^2} \nonumber \\
&= - \frac{\partial}{\partial x[j]} \sum_{i=0}^{L-1} \frac{( y_m[i] - (Z x)[(i+l)~\mathrm{mod}~2L] )^2}{2\sigma^2} \nonumber \\
&= - \frac{1}{\sigma^2} \sum_{i=0}^{L-1} \bigg( y_m[i] - (Z x)[(i+l)~\mathrm{mod}~2L] \bigg) \nonumber \\
&\hspace{6em} \times \delta[(i+l) - (j+L)].
\end{align}
\endgroup
Substituting this expression into (\ref{xderv_L}), we obtain
\begin{align}
\sum_{m=0}^{N_d-1} \sum_{l=j+1}^{j+L} p(l | y_m, x_k) (y_m[j+L-l] - x[j]) = 0. \nonumber
\end{align}
Rearranging the terms gives the update rule shown in (\ref{update_single1}).

In order to update the priors, we maximize $\mathcal{L}(x, \alpha, \lambda)$ with respect to $\alpha$:
\begin{align}
0 = \frac{\partial \mathcal{L}}{\partial \alpha[l]} = \frac{1}{\alpha[l]} \sum_{m=0}^{N_d-1} p(l | y_m, x_k) - \lambda, \nonumber
\end{align}
where $0 \leq l < 2L$.
We thus obtain the update rule for $\alpha$ as
\begin{align}
\alpha[l] = \frac{1}{\lambda} \sum_{m=0}^{N_d-1} p(l | y_m, x_k), \nonumber
\end{align}
and we can immediately solve $\lambda = N_d$ from the normalization $\sum_{l=0}^{2L-1} \alpha[l] = 1$.

\section{Derivation of \eqref{update_double}}
\label{derv_update_double}
As shown in (\ref{Q_double}), $Q(x, \alpha | x_k, \alpha_k)$ is additively separable for $x$ and $\alpha$.
Moreover, the constraint \eqref{constraint} only involves the values of the parameters $\alpha[0], \alpha[1], \rho_1$, so the maximization of $Q(x, \alpha | x_k, \alpha_k)$ is unconstrained on $x$.
At the maximum of $Q(x, \alpha | x_k, \alpha_k)$, we have
\begingroup
\allowdisplaybreaks
\begin{align}
\label{xderv_Q}
&\frac{\partial Q(x, \alpha | x_k, \alpha_k)}{\partial x[j]} = 0 \nonumber \\
&= \sum_{m=0}^{N_d-1} \bigg[ \sum_{l=0}^{2L-1} p(l | y_m, x_k) \frac{\partial \log p(y_m | l, x)}{\partial x[j]} \nonumber \\
&\hspace{6pt} + \sum_{l_1=L+1}^{2L-1} \sum_{l_2=1}^{l_1-L} p(l_1, l_2 | y_m, x_k) \frac{\partial \log p(y_m | l_1, l_2, x)}{\partial x[j]} \bigg],
\end{align}
\endgroup
where $0 \leq j < L$.
With $p(y_m | l_1, l_2, x_k)$ given by \eqref{prob_double}, we obtain
\begin{align}
\label{xderv_double}
&\frac{\partial \log p(y_m | l_1, l_2, x)}{\partial x[j]} \nonumber \\
&= - \frac{\partial}{\partial x[j]} \sum_{i=0}^{L-1} \frac{( y_m[i] - (C R_{l_1} Z x)[i] - (C R_{l_2} Z x)[i] )^2}{2\sigma^2} \nonumber \\
&= - \frac{1}{2\sigma^2} \frac{\partial}{\partial x[j]} \sum_{i=0}^{L-1} \bigg( y_m[i] - (Z x)[(i+l_1)~\mathrm{mod}~2L] \nonumber \\
&\hspace{1em} - (Z x)[(i+l_2)~\mathrm{mod}~2L] \bigg)^2 \nonumber \\
&= - \frac{1}{\sigma^2} \sum_{i=0}^{L-1} \bigg( y_m[i] - (Z x)[(i+l_1)~\mathrm{mod}~2L] \nonumber \\
&\hspace{1em} - (Z x)[(i+l_2)~\mathrm{mod}~2L] \bigg) \nonumber \\
&\hspace{1em}\times \bigg( \delta[(i+l_1) - (j+L)] + \delta[(i+l_2) - (j+L)] \bigg).
\end{align}
Substituting (\ref{xderv_single}) and (\ref{xderv_double}) into (\ref{xderv_Q}), we obtain
\begingroup
\allowdisplaybreaks
\begin{align}
0 &= \sum_{m=0}^{N_d-1} \bigg[ \sum_{l=j+1}^{j+L} p(l | y_m, x_k) (y_m[j+L-l] - x[j]) \nonumber \\
	&+ \sum_{l_1=L+1}^{j+L} \sum_{l_2=1}^{l_1-L} p(l_1, l_2 | y_m, x_k) (y_m[j+L-l_1] - x[j]) \nonumber \\
	&+ \sum_{l_1=L+j+1}^{2L-1} \sum_{l_2=j+1}^{l_1-L} p(l_1, l_2 | y_m, x_k) (y_m[j+L-l_2] - x[j]) \bigg]. \nonumber
\end{align}
\endgroup
Rearranging the terms, we obtain the update rule in (\ref{update_double}).

\ifCLASSOPTIONcaptionsoff
  \newpage
\fi

\end{document}